\documentclass[twocolumn]{aastex631}

\shorttitle{AASTeX v6.3.1 Sample article}
\shortauthors{Sarkar et al.}
\graphicspath{{./}{figures/}}
\begin{document}

\title{Modelling the magnetic vectors of ICMEs at different heliocentric distances with INFROS}

\correspondingauthor{Ranadeep Sarkar}
\email{ranadeep.sarkar@helsinki.fi}

\author{Ranadeep Sarkar}
\affiliation{Department of Physics, University of Helsinki, P.O. Box 64, FI-00014 Helsinki, Finland}

\author{Nandita Srivastava}
\affiliation{Udaipur Solar Observatory, Physical Research Laboratory, Udaipur 313001, India}

\author{Nat Gopalswamy}
\affiliation{NASA Goddard Space Flight Center, Greenbelt, MD 20771, USA}

\author{Emilia Kilpua}
\affiliation{Department of Physics, University of Helsinki, P.O. Box 64, FI-00014 Helsinki, Finland}

\begin{abstract}
The INterplanetary Flux ROpe Simulator (INFROS) is an observationally constrained analytical model dedicated for forecasting the strength of the southward component (Bz) of the magnetic field embedded in interplanetary coronal mass ejections (ICMEs). In this work, we validate the model for six ICMEs  sequentially observed by two radially-aligned spacecraft positioned at different heliocentric distances. The six selected ICMEs in this study comprise of cases associated with isolated CME evolution as well as those interacting with high-speed streams (HSS) and high-density streams (HDS). For the isolated CMEs, our results show that the model outputs at both the spacecraft are in good agreement with in-situ observations. However, for most of the interacting events, the model correctly captures the CME evolution only at the inner spacecraft. Due to the interaction with HSS and HDS, which in most cases occurred at heliocentric distances beyond the inner spacecraft, the ICME evolution no longer remains self-similar. Consequently, the model underestimates the field strength at the outer spacecraft. Our findings indicate that constraining the INFROS model with inner spacecraft observations significantly enhances the prediction accuracy at the outer spacecraft for the three events undergoing self-similar expansion, achieving a 90\% correlation between observed and predicted Bz profiles. This work also presents a quantitative estimation  of the ICME magnetic field enhancement due to interaction which may lead to severe space weather. We conclude that the assumption of self-similar expansion provides a lower limit to the magnetic field strength estimated at any heliocentric distance, based on the remote sensing observations.

\end{abstract}
\section{Introduction}
One of the key challenges in space-weather forecasting is to reliably determine the strength and orientation of the magnetic field inside an Earth-impacting ICME in near-real time. Several efforts have been made to forecast the magnetic vectors of ICMEs using various analytical and numerical models \citep{odstr_pizzo,shen_2014,Savani_2015,kay,Mostl2018,2018Pomoell,2019scoloni,kilpua_2019,Sarkar_2020,2022pal, Sarkar_2024}. Efforts are still on-going to improve such models as they have not yet been able to accurately predict the southward component (Bz) of CME magnetic field based on their validation results.

\begin{figure*}[tp]
\begin{center}
\centering
\includegraphics[width=0.4\textwidth,clip=]{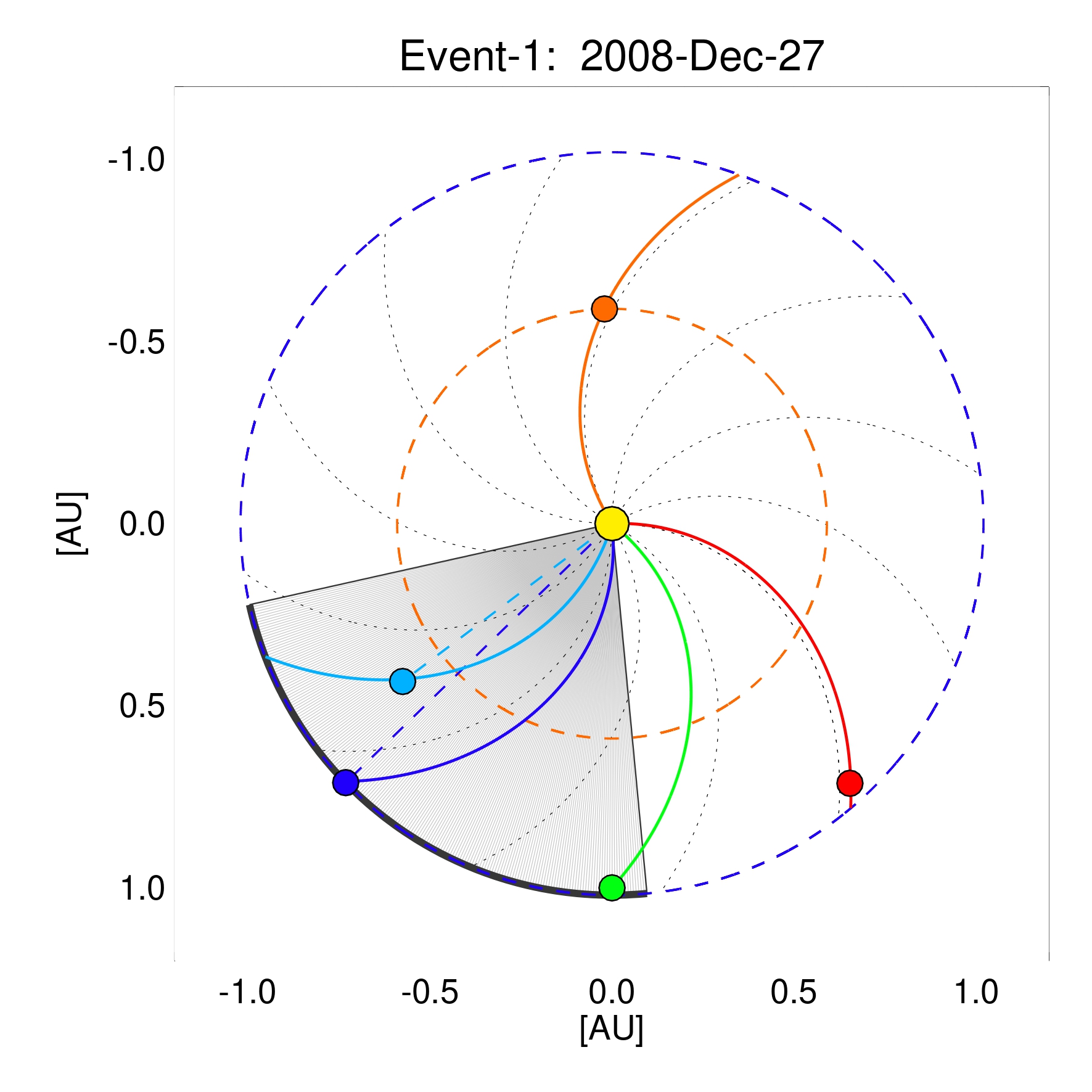}
\includegraphics[width=0.4\textwidth,clip=]{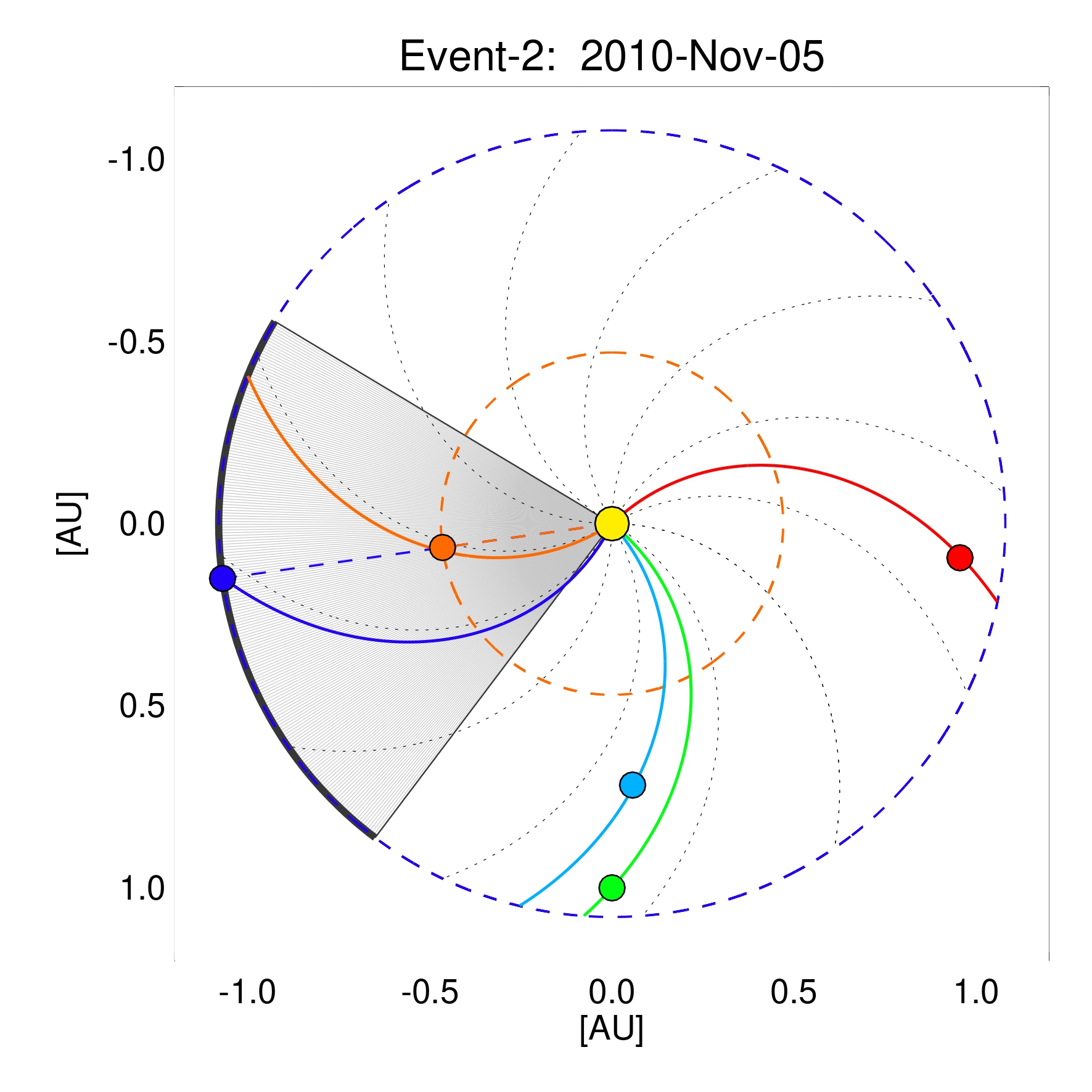}
\includegraphics[width=0.4\textwidth,clip=]{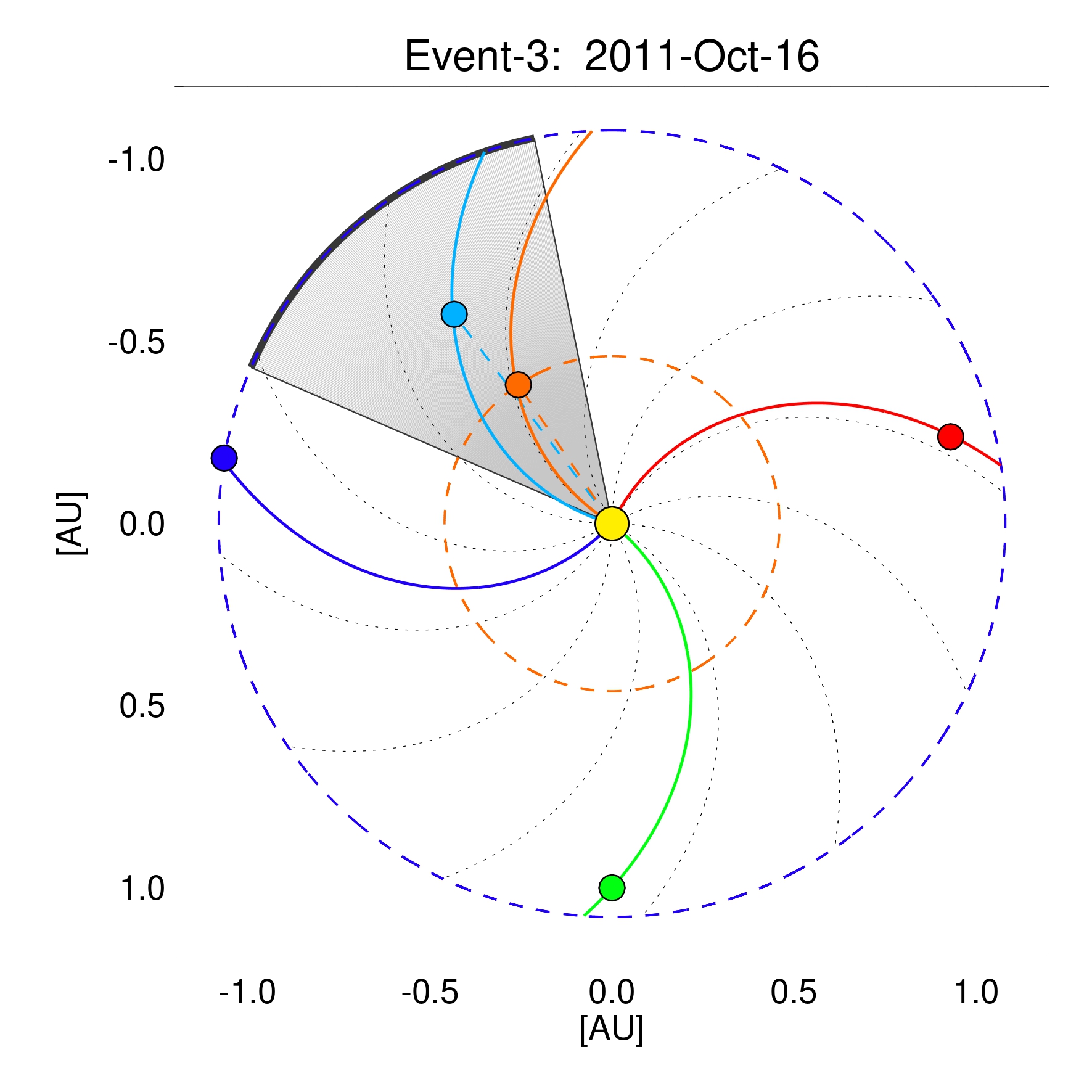}
\includegraphics[width=0.4\textwidth,clip=]{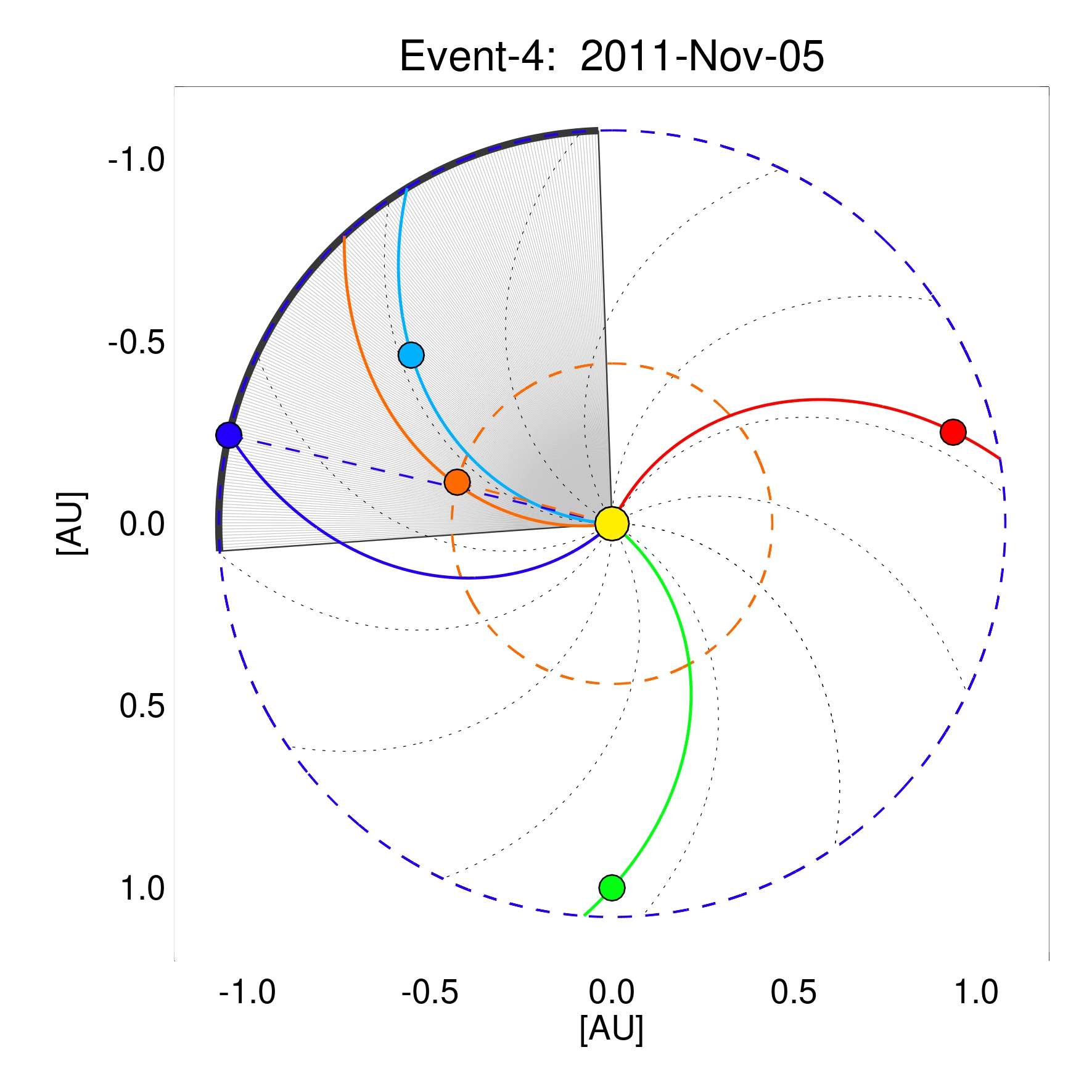}
\includegraphics[width=0.4\textwidth,clip=]{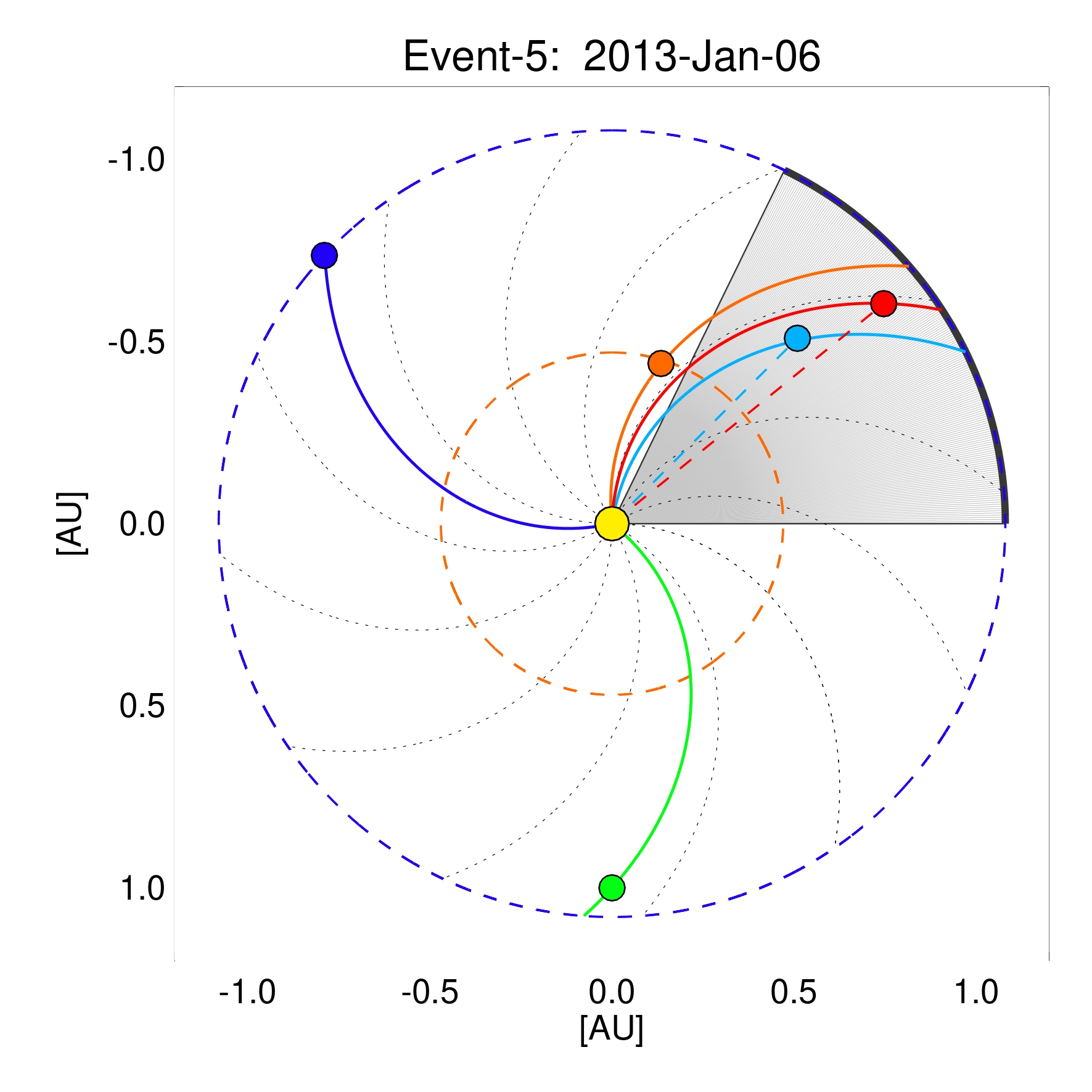}
\includegraphics[width=0.4\textwidth,clip=]{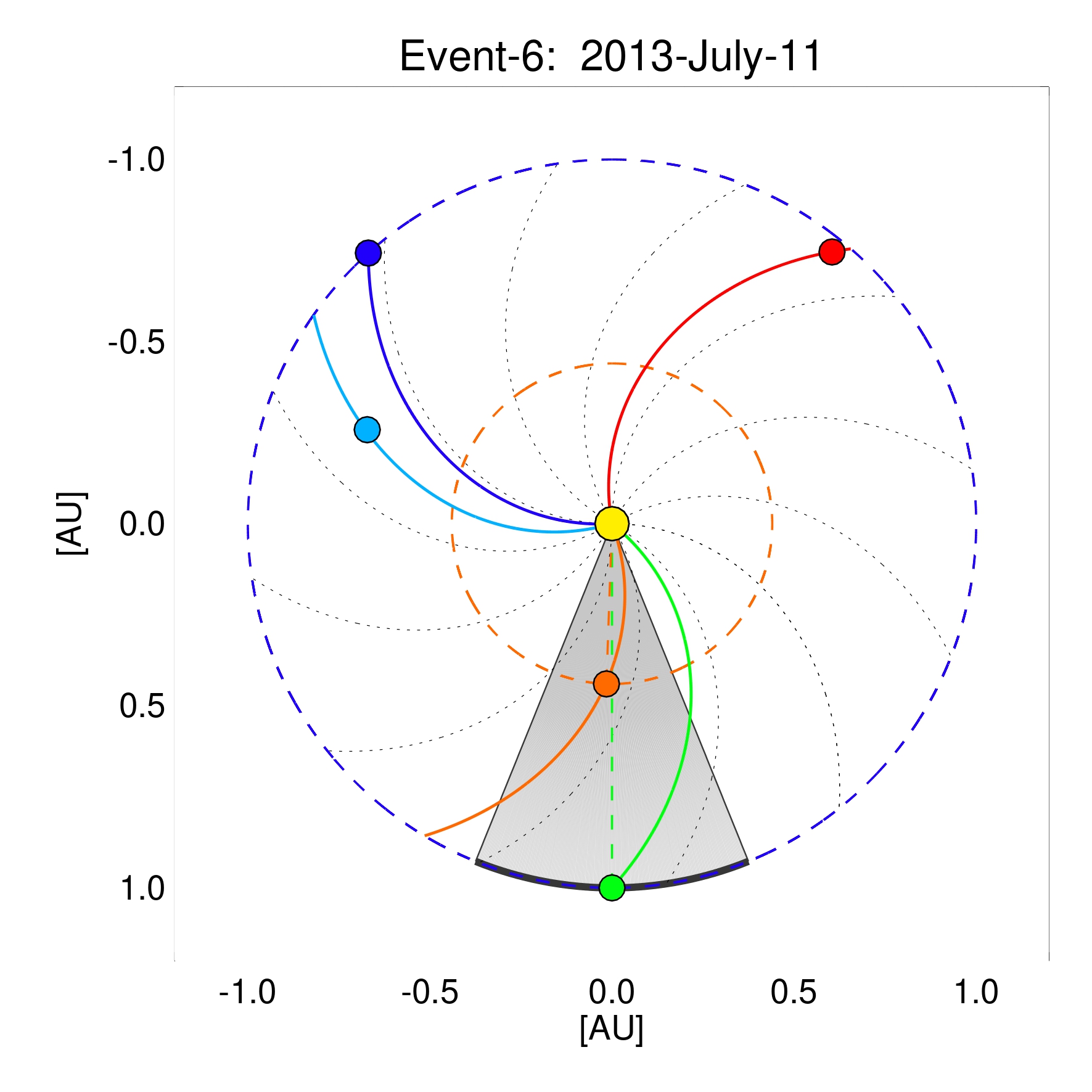}
\includegraphics[width=0.9\textwidth,clip=]{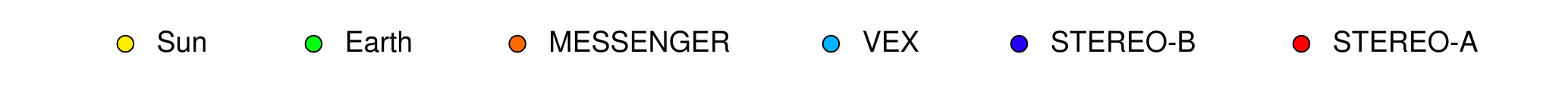}
\end{center}
\caption{Location of MESSENGER, VEX, STEREO and Earth in the ecliptic plane during the ICME events under study. The gray shaded regions denote the propagation path of the ICMEs. The outer boundary of each plot is set at the location of STEREO-B. The Parker spirals are drawn assuming a solar wind speed of 450 $km\ s^{-1}$.\label{where_is_stereo}}
\end{figure*}

The physics-based global MHD models can capture the detailed evolution of a CME incorporating its interaction with the background solar wind or other CMEs. However, these tools are computationally expensive and difficult to run in real time for the magnetised CME models. On the other hand, analytical modelling tools are based on semi-empirical methods and can be run in real-time for operational space-weather prediction purposes. The analytical modelling framework as developed in the INterplanetary Flux ROpe Simulator (INFROS) \citep{Sarkar_2020}, serves as a tool to estimate the magnetic vectors of ICMEs in near-real time. 

INFROS has shown remarkably good results in its initial validation with an observed Earth-impacting CME \citep{Sarkar_2020}. Notably, INFROS assumes the CME evolution to be self-similar, which is overall valid for the isolated CME evolution in the absence of any interaction with other CMEs or solar wind structures such as high-speed streams (HSS) and high-density streams (HDS). Therefore, it is important to understand how  this model works in the case of CMEs interacting with other solar wind structures. Moreover, the earlier validation of INFROS was performed for a CME observed at a single vantage point at 1 AU. Therefore, it would be an important step forward to validate the model for both interacting and non-interacting CMEs observed sequentially by two radially-aligned spacecraft at different heliocentric distances. This approach of validating the model allows us to study a comparable part of an ICME at different radial distances as it propagates away from the Sun. Furthermore, by including both interacting and non-interacting ICME events in our study, we can gain valuable insights into how different solar wind conditions influence the evolution of ICMEs. Importantly, this approach will provide important clues about how the evolution of the ICME differs from self-similar expansion in cases where interactions occur.

\begin{deluxetable*}{cccccccccc}[!t]
\tablecaption{ICME events sequentially detected by two radially aligned spacecraft positioned at different heliocentric distances R1 (inner) and R2 (outer) respectively. The latitudinal and longitudinal separations between the pair of spacecraft are given by $\Delta \theta$ and $\Delta \phi$ respectively.  \label{table1}}
\tablehead{
\colhead{Event} & \colhead{Spacecraft 1} &\colhead{R1} & \colhead{Spacecraft 2} & \colhead{R2} & \colhead{$\Delta \theta$} & \colhead{$\Delta \phi$} &\colhead{Arrival time of  ICME} & \colhead{Arrival time of ICME} & \colhead{ICME}\\
\colhead{no} & \colhead{ } &\colhead{[AU]} & \colhead{ } & \colhead{[AU]} & & & \colhead{in Spacecraft 1} & \colhead{in Spacecraft 2} & \colhead{interacting}\\
\colhead{} & \colhead{} & \colhead{} & \colhead{} & \colhead{} & & &\colhead{yyyy/mm/dd hh:mm} & \colhead{yyyy/mm/dd hh:mm} & \colhead{with}
}

\startdata
1 & VEX & 0.72 & STEREO-B  & 1.03  & 1.1$^{\circ}$& 8.8$^{\circ}$&2008/12/29 20:46 UT& 2008/12/31 03:39 UT & HSS \\
2 & MESSENGER & 0.45 & STEREO-B & 1.08   & 7.0$^{\circ}$& 1.3$^{\circ}$&2010/11/05 11:46 UT & 2010/11/07 19:05 UT & HSS\\
3 & MESSENGER & 0.46 & VEX  & 0.72  & 2.6$^{\circ}$& 2.0$^{\circ}$&2011/10/15 08:25 UT& 2011/10/16 00:50 UT & none\\
4 & MESSENGER & 0.44 & STEREO-B & 1.08  & 6.8$^{\circ}$& 4.3$^{\circ}$&2011/11/04 15:09 UT& 2011/11/06 05:10 UT & none\\
5 & VEX & 0.73 & STEREO-A  & 0.96  & 1.4$^{\circ}$& 5.5$^{\circ}$&2013/01/08 09:22 UT & 2013/01/09 02:25 UT & HSS\\
6 & MESSENGER & 0.45 &  Wind & 1.00  & 6.0$^{\circ}$& 0.9$^{\circ}$&2013/07/11 00:00 UT & 2013/07/13 04:30 UT & HDS\\
\enddata
\end{deluxetable*}

This article is organized as follows. First, we briefly discuss the model architecture of INFROS in Section \ref{model_description} and the selection criteria for ICMEs observed by multi spacecraft in Section \ref{event_slection}. Estimation of the near-Sun flux rope (FR) parameters for the selected events are described in Section \ref{sec_input}. The INFROS model results for the six selected events are presented in Section \ref{results}. Finally, we conclude our results in Section \ref{summary}.

\begin{figure*}[t!]
\begin{center}
\centering
\includegraphics[width=.96\textwidth,clip=]{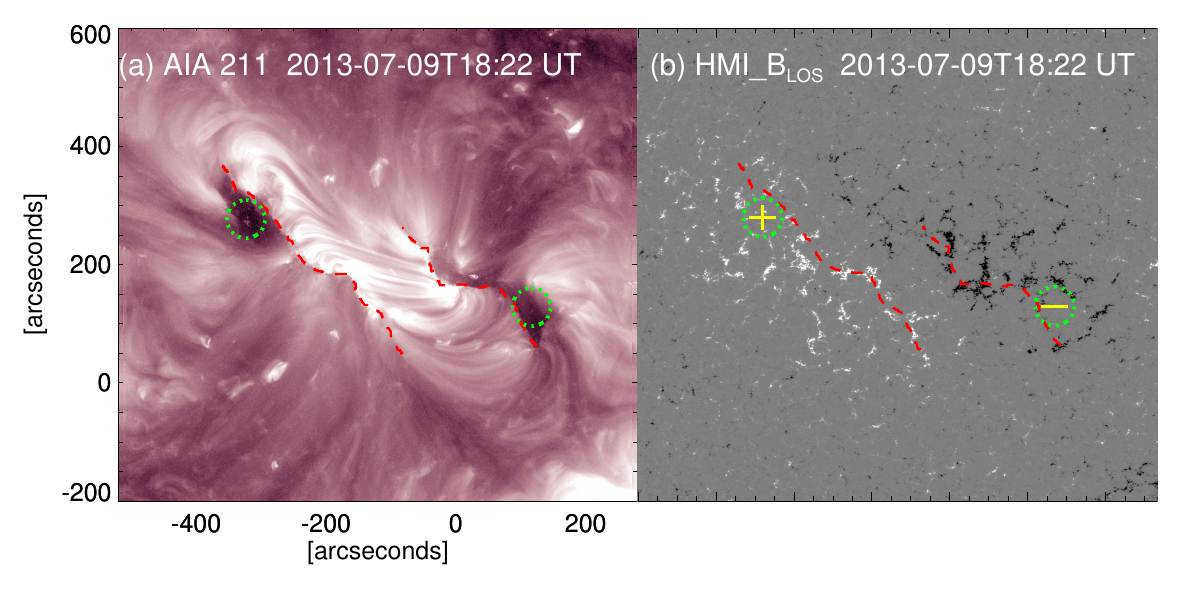}
\end{center}
\caption{Observations for the source region of Event-6. (a) Depicts the morphology of the PEA as observed in AIA 211 \AA\ image. The red dashed lines mark the boundary of the PEA region. The green dotted circles denote the location of twin core dimming regions.  (b) Illustrates the line-of-sight component of HMI magnetic field. The same red dashed boundary and the green dotted circles as shown in (a) are over-plotted in panel (b). The underlying magnetic polarities of the core dimming locations as denoted by the green dotted circles are shown which indicate a south-west directed left-handed FR.}\label{EVENT_6_source_region}
\end{figure*}

\section{Model description}\label{model_description}
INFROS is an observationally constrained analytical modelling framework, that models the evolution of magnetic vectors of ICMEs in interplanetary space \citep{Sarkar_2020}. It uses the near-Sun observations to constrain the magnetic and geometric properties of a CME FR observed close ($\geq 10$ solar radii) to Sun. The detailed input parameters required to constrain the model are discussed in Section \ref{sec_input}. Once the tilt and direction of propagation of the FR is constrained from near-Sun observations at or above 10 solar radii ($R_S$), the model assumes that there is no further rotation and deflection of the ICME throughout its rest of the propagation path. Based on this assumption the impact distance (closest distance to the magnetic axis of the flux rope from the location of the spacecraft) of the ICME at any targeted spacecraft is determined. Further, the evolution of the size and the axial magnetic field strength of the FR at any heliocentric distance is modelled using the assumption of self-similar expansion and the conservation of magnetic flux inside the FR.  The final magnetic vector profiles of the ICME along its passage through the targeted spacecraft is obtained  using a cylindrical flux rope solution \citep{lundquist1950magnetohydrostatic}. In contrast to existing models (e.g.\citet{Savani_2015,kay,Mostl2018}), INFROS is formulated in such a way that it does not involve any free parameters like the dimension, axial field strength, time of passage and the speed of ICME to model its expanding nature as it passes through the in situ spacecraft. All the input parameters of this model are constrained from the remote-sensing observations. A detailed description of the model is given in \citet{Sarkar_2020}.      

\section{Event selection}\label{event_slection}
Earlier studies have reported on the ICME events observed by the radially-aligned multiple spacecraft at different heliocentric distances \citep{Mostl2018,2019JGRA..124.4960G,Weiss_2021}. Out of those events, we have selected the ones which satisfy the following criteria 

\begin{itemize}
\item The longitudinal separation between the two spacecraft should be less than 10$^\circ$ so that both the spacecraft encounter nearly the same part of the ICME.
\item The ICMEs should contain the signatures of a magnetic cloud at both spacecraft so that the magnetic vectors of the ICME can be associated with FR signatures.
\item There should not be interaction between two CMEs as inclusion of CME-CME interaction is beyond the scope of the analytical modelling framework of INFROS.

\item CME source-region signatures should be detected from on-disk observations either by SDO or STEREO including the far-side events in order to estimate some of the magnetic properties (chirality and axis-orientation) of the CMEs, thereby excluding the events associated with stealth CMEs. 
\end{itemize}
Based on the above mentioned criteria, we have selected six events for the present study as listed in Table \ref{table1}. The table also provides information on the type of ICME events, categorizing them as isolated or interacting cases. The locations of the different spacecraft e.g. MErcury Surface, Space ENvironment,
GEochemistry, and Ranging (MESSENGER) \citep{MESSENGER}, Venus Express (VEX) \citep{VEX}, Solar TErrestrial RElations Observatory (STEREO) \citep{STEREO} and Wind \citep{Wind} during those selected ICME events are shown in Figure \ref{where_is_stereo}. The insitu data from MESSENGER, STEREO and Wind are obtained from CDAWeb\footnote{\url{https://cdaweb.gsfc.nasa.gov/}}, and the data from VEX are obtained from ESA's Planetary Science Archive\footnote{\url{https://archives.esac.esa.int/psa/ftp/VENUS-EXPRESS/MAG/}}.

\begin{deluxetable*}{ccccc}[!t]
\tablecaption{Information on CMEs associated with the ICME events\label{table_CME}}
\tablehead{
\colhead{Event no} & \colhead{CME observation time} & \colhead{Apex height from GCS} & \colhead{Speed from GCS} & \colhead{Poloidal flux} \\
\colhead{} & \colhead{yyyy/mm/dd hh:mm}  & \colhead{$R_S$} & \colhead{km s$^{-1}$} & \colhead{$\times 10^{21}$ Mx}
}

\startdata
1 & 2008-12-27 10:53 UT & 12 & 570 & 1.7\\
2 & 2010-11-04 01:54 UT & 16.5 & 808 & 2.9\\
3 & 2011-10-14 13:54 UT & 10 & 920 & 3.5\\
4 & 2011-11-03 23:54 UT & 10 & 760 & 2.7\\
5 & 2013-01-06 08:54 UT & 11.2 & 580 & 1.8 \\
6 & 2013-07-09 17:54 UT & 11 & 600 & 1.6 \\
\enddata
\end{deluxetable*}

\begin{figure*}[t!]
\begin{center}
\centering
\includegraphics[width=\textwidth,clip=]{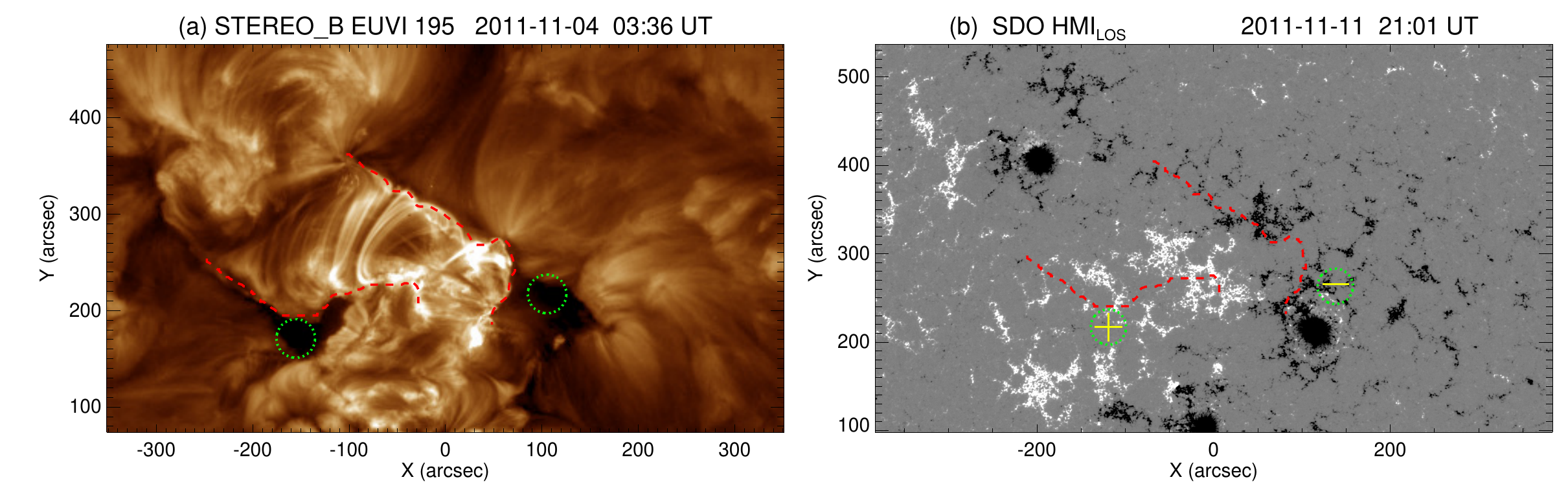}
\end{center}
\caption{Observations of the source region for Event-4. (a) Depicts the morphology of the PEA as observed in STEREO-B EUVI 195 \AA\ image. The red dashed lines mark the boundary of the PEA region. The green dotted circles denote the location of twin core dimming regions.  (b) Illustrates the line-of-sight component of HMI magnetic field within the same field-of-view as (b). (a) and (b) are approximately co-spatial as the patch center of both images belongs to the same Carrington longitude and latitude. The red dashed boundary and the green dotted circles as shown in (a) are over-plotted in panel (b) using the information on their respective Carrington longitude and latitude. The underlying magnetic polarities of the core dimming locations as denoted by the green dotted circles are shown which indicate a westward directed left-handed FR.}\label{st_hmi}
\end{figure*}

\begin{deluxetable*}{cccccccc}[!t]
\tablecaption{Geometric and magnetic parameters of the CME FR at 10 R$_s$: Inputs for the INFROS model\label{table2}}
\tablehead{
\colhead{Event no} & \colhead{Chirality} &\colhead{Axial field direction} & \colhead{Axial field strength} & \colhead{Tilt angle} & \colhead{Aspect ratio} & \colhead{Longitude} & \colhead{Latitude}\\
\colhead{} & \colhead{ } &\colhead{ } & \colhead{mG} &  &\colhead{$\kappa$} & \colhead{in HEEQ} & \colhead{in HEEQ}
}
\startdata
1 & Left handed & Westward & 30 $\pm$ 05 & -5$^{\circ}$ $\pm$ 10$^{\circ}$ & 0.26 $\pm$ 0.02 & -36$^{\circ}$ $\pm$ 10$^{\circ}$& 14$^{\circ}$ $\pm$ 10$^{\circ}$\\
2 & Right handed & Eastward & 58 $\pm$ 05 & 21$^{\circ}$ $\pm$ 10$^{\circ}$ & 0.23 $\pm$ 0.02 & -79$^{\circ}$ $\pm$ 10$^{\circ}$ & 0$^{\circ}$ $\pm$ 10$^{\circ}$\\
3 & Left handed & Southward & 68 $\pm$ 05 & 70$^{\circ}$ $\pm$ 10$^{\circ}$ & 0.24 $\pm$ 0.02 & -140$^{\circ}$ $\pm$ 10$^{\circ}$ & 6$^{\circ}$ $\pm$ 10$^{\circ}$\\
4 & Left handed & Westward & 40 $\pm$ 05 & -35$^{\circ}$ $\pm$ 10$^{\circ}$ & 0.26 $\pm$ 0.02 & -134$^{\circ}$ $\pm$ 10$^{\circ}$ & -2$^{\circ}$ $\pm$ 10$^{\circ}$\\
5 & Right handed & Northward & 32 $\pm$ 05 & 45$^{\circ}$ $\pm$ 10$^{\circ}$ & 0.26 $\pm$ 0.02 & 122$^{\circ}$ $\pm$ 10$^{\circ}$ & -13$^{\circ}$ $\pm$ 10$^{\circ}$\\
6 & Left handed & Westward & 40 $\pm$ 05 & -5$^{\circ}$ $\pm$ 10$^{\circ}$ & 0.28 $\pm$ 0.02 & -6$^{\circ}$ $\pm$ 10$^{\circ}$ & 3$^{\circ}$ $\pm$ 10$^{\circ}$\\
\enddata

%\tablecomments{}
\end{deluxetable*}
\begin{figure*}[t!]
\begin{center}
\centering
\includegraphics[width=\textwidth,clip=]{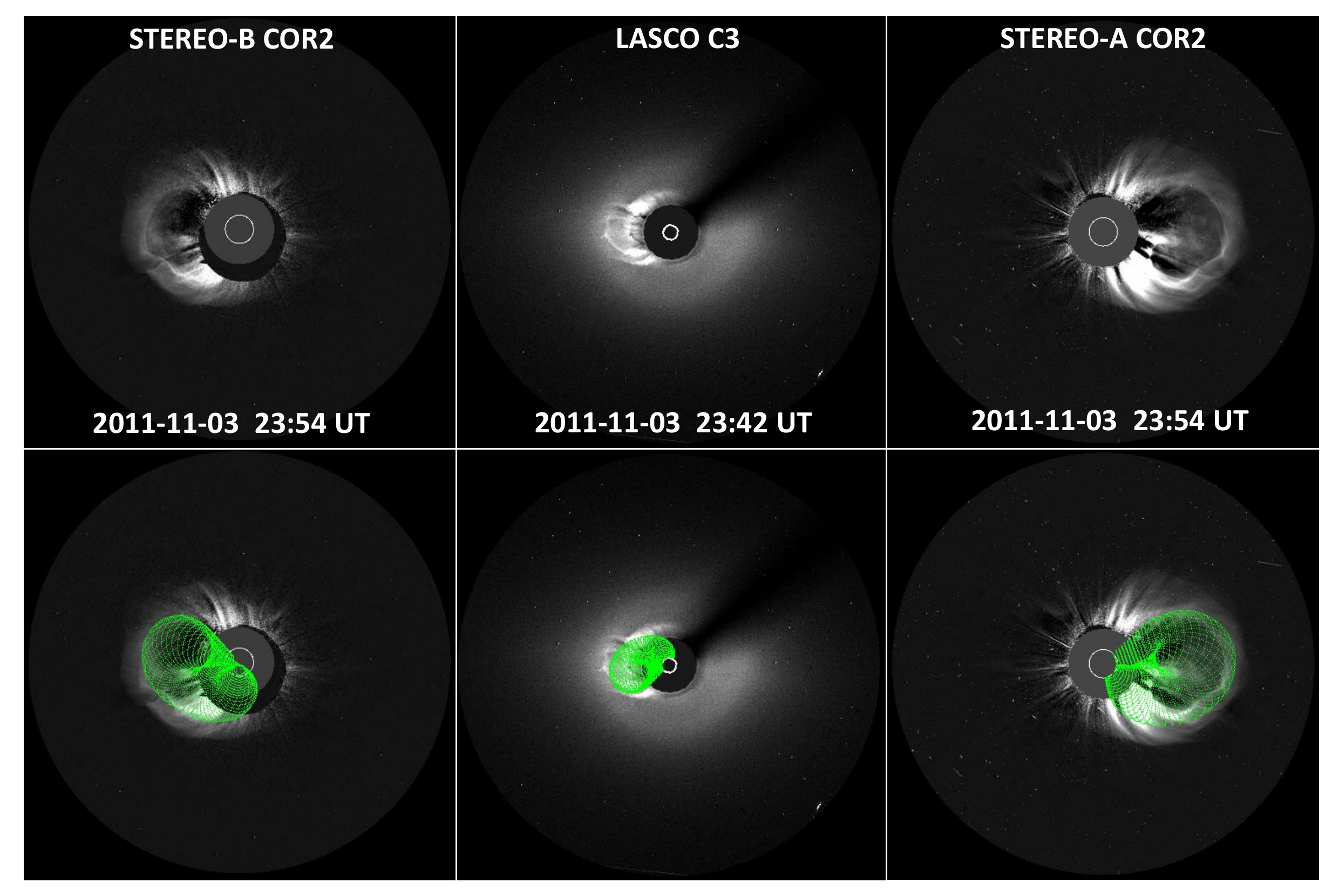}
\end{center}
\caption{Top panels depict the CME morphology observed in COR2-B (top-left), LASCO C3 (top-middle), and COR2-A (top-right), respectively for Event-4. Bottom panels illustrate the overplot of the best-fitted wire frame (green dotted marks) of the FR using the GCS model.}\label{gcs_pic}
\end{figure*}

\section{Observationally constrained input parameters}\label{sec_input}

INFROS uses  observational inputs to constrain the magnetic, geometric and kinematic properties of a near-Sun FR. Therefore, we investigate the near-Sun CME properties of our ICME events.  Contextually, the magnetic parameters of a FR refer to its chirality, axial direction and the magnetic flux which can be estimated using the EUV observations of the erupting source region on the Sun, combined with the information on the underlying photospheric magnetic field \citep{Palmerio2017,2018Gopal,Sarkar_2020}. On the other hand, the geometric and kinematic parameters include the tilt angle, edge-on and face-on extent of the FR along with the speed and direction of propagation. These parameters can be obtained by applying a graduated cylindrical shell (GCS) model \citep{gcs2,gcs1} to the multi-vantage point white-light observations of the CME morphology.       

Combining the white-light coronagraphic observations obtained by SOHO and STEREO spacecraft, we first identify the CME eruptions that are associated with the ICME events under study. We apply the GCS model to those identified CMEs to estimate their 3D speed and the other associated geometric properties. Table \ref{table_CME} lists the observation time of all the CMEs and their corresponding apex height during the time of observations and the associated 3D speed as obtained from GCS. Next, we investigate the source region characteristics of those CMEs to estimate the associated magnetic properties. 

Earlier studies indicate that the poloidal flux of magnetic flux ropes generated by reconnection is roughly equivalent to the reconnection flux in the low corona \citep{Qiu2007,2007Longcope}. This can be estimated either through the photospheric magnetic flux beneath the combined area covered by the flare ribbons \citep{Kazachenko_2017}, or by examining the magnetic flux beneath the post-eruption arcades (PEAs) \citep{gopalswamy2017,gopalswamy2018}. However, we notice that the source region of all but one CME is located either at the back side of the Sun or close to the solar limb with respect to Earth. Therefore, it was not possible to perform such analysis of the poloidal flux for the back-sided events due to the absence of far-side photospheric magnetic field data and lack of reliability of the magnetogram data for the limb events. Considering this fact, we use the empirical relation between the 3d speed of a CME and the associated reconnection flux to estimate the poloidal flux of the CME FR \citep{2018Gopal} for the first five events as listed in Table \ref{table_CME}. 

As the source region of Event-6 was located close to the solar disk-center as viewed from Earth, we obtain the  poloidal flux of this event using the PEA as captured in Atmospheric Imaging Assembly (AIA; \citeauthor{AIA} \citeyear{AIA}) 211~\AA\ images and the magnetic field data from Helioseismic and Magnetic Imager (HMI;  \citeauthor{HMI} \citeyear{HMI}) on board the Solar Dynamics Observatory (SDO; \citeauthor{SDO} \citeyear{SDO}). Figure \ref{EVENT_6_source_region} depicts the morphology of the PEA as formed during the eruption. The magnetic flux underlying this PEA region yields a value of $1.6 \times 10^{21}$ Mx, which determines the poloidal flux of the erupting FR. This value is consistent with the one obtained from the empirical relation:$1.8 \times 10^{21} $ Mx. We list the values of poloidal flux for all the events in Table \ref{table_CME}.  

In order to estimate the axial field direction of the associated FRs, we first identify the core dimming regions observed during their eruption phase. The green dotted circles drawn on Figure \ref{EVENT_6_source_region} depict the conjugate core dimming locations identified for the Event-6. The underlying magnetic polarities of these core-dimming locations (see Panel (b) of Figure \ref{EVENT_6_source_region}) clearly indicate that the axial-field of the associated FR points towards south-west direction. Further, the magnetic connectivity between the two core-dimming regions refers to a left-handed chirality of the flux-rope. This negative chirality sign can also be confirmed from the observed reverse S-shaped (northern hemisphere) morphology of the post-eruptive loops or the reverse J-shaped boundary of the core-dimming regions.    

\begin{figure*}[h!]
\begin{center}
\includegraphics[width=0.75\textwidth,clip=]{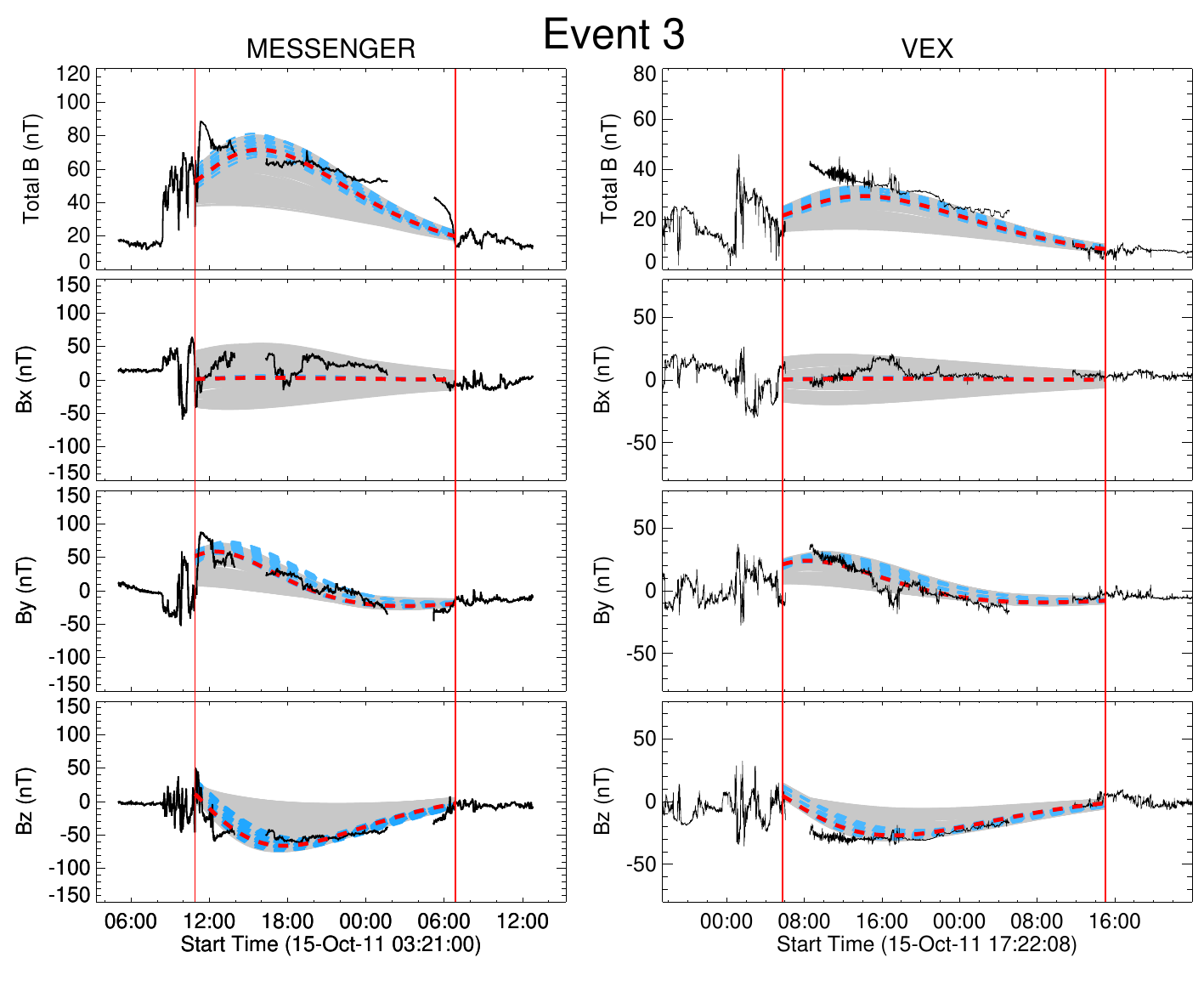}
\includegraphics[width=0.75\textwidth,clip=]{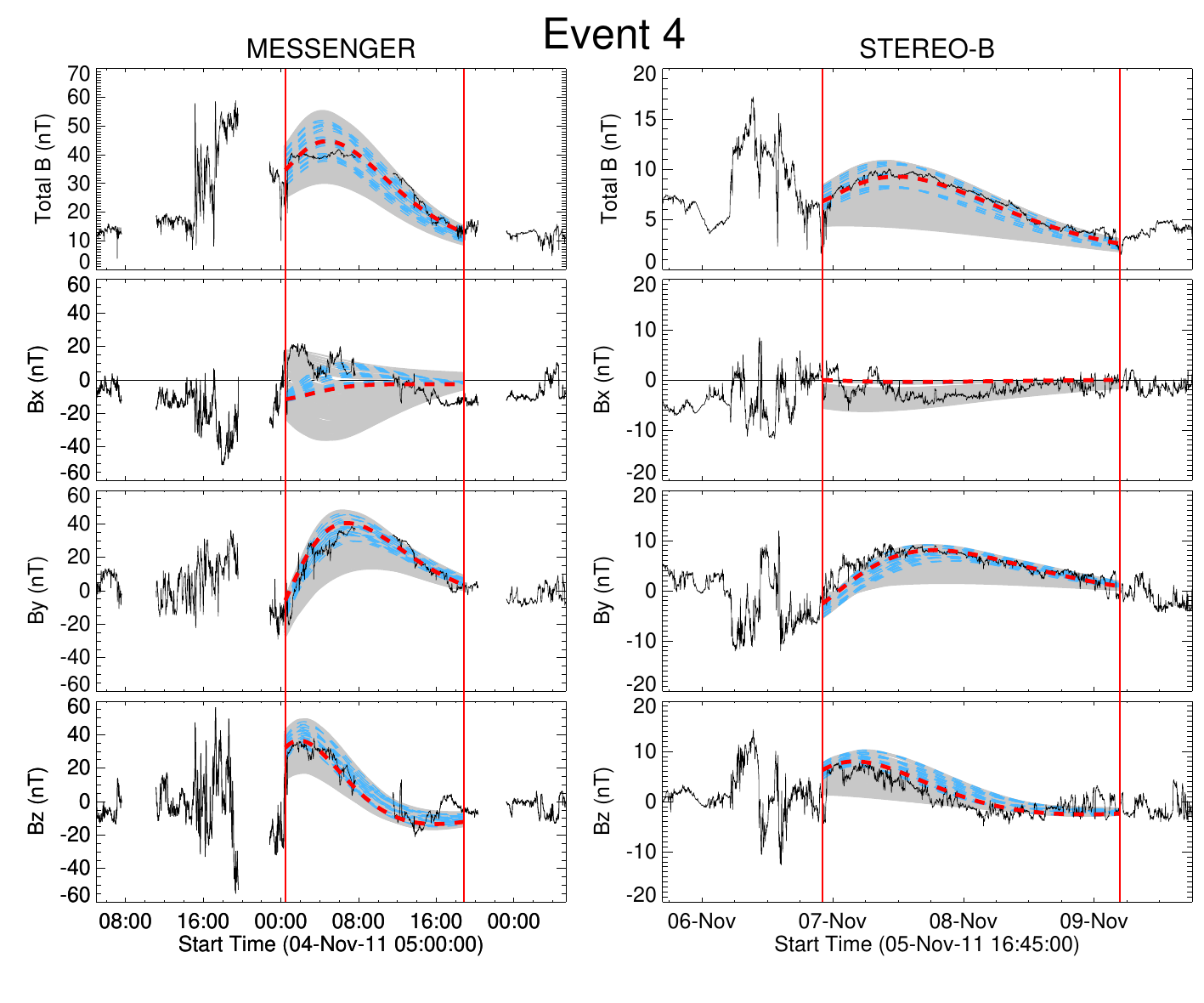}
\caption{Top panels show the B‐field data (in SCEQ coordinates) of the ICME sequentially detected by MESSENGER and VEX for Event-3. Bottom panels depict the same for the ICME sequentially detected by MESSENGER and STEREO-B for Event-4. The black solid lines within the boundary marked by the vertical red lines, denote the magnetic vectors of the ICME. The gray shaded regions are the ensemble results obtained from the model predictions. The cyan dashed lines represent the ensemble model results corresponding to the lowest impact distance at the outer spacecraft. Amongst these ensemble model results, the one that best fits the magnetic vectors observed by MESSENGER is denoted as a red dashed line in each plot.\label{ICME3_4}}
\end{center}
\end{figure*}

For the back-sided and limb events (first five events in Table \ref{table_CME}), we investigate the location of core dimmings, PEA morphology or the S-shaped sigmoidal structure using the observation from STEREO EUVI \citep{Wuelser2004}. In order to get the information on the underlying photospheric magnetic field of those source regions, we use the co-spatial HMI observations taken after a few days when they appear on the visible solar disk as observed from Earth. The core dimming locations as detected during the CME eruptions from STEREO observations are then overlaid on the HMI magnetogram using the information on the Carrington longitude and latitude of the source region of the CMEs (Figure \ref{st_hmi}). Thus, we obtain the axial direction as well as the chirality of the associated magnetic FRs for all the events under study.

Using the information on the poloidal flux as listed in Table \ref{table_CME} and the geometric parameters of the CMEs as obtained from the GCS reconstruction (see Figure \ref{gcs_pic}), we apply the FRED (Flux Rope from Eruption Data) \citep{gopalswamy2018} model to estimate the axial field strength ($B_{0}$) of the CMEs at 10 $R_s$ (Table \ref{table2}). It is important to note that for the majority of events, the poloidal flux is estimated based on its empirical relation with the 3D speed of a CME. Consequently, we consider an uncertainty of $\pm 5$ mG while estimating $B_{0}$ to account for the associated uncertainty of approximately $\pm$ 100 km s$^{-1}$ in the estimated CME speed. The range of uncertainties for the other geometric parameters of the CMEs is chosen following the study presented in \citet{Sarkar_2020}. All the magnetic and geometric parameters of the CMEs, along with their uncertainties are listed in Table \ref{table2}.

\begin{figure*}[btp]
\begin{center}
\includegraphics[width=\textwidth,clip=]{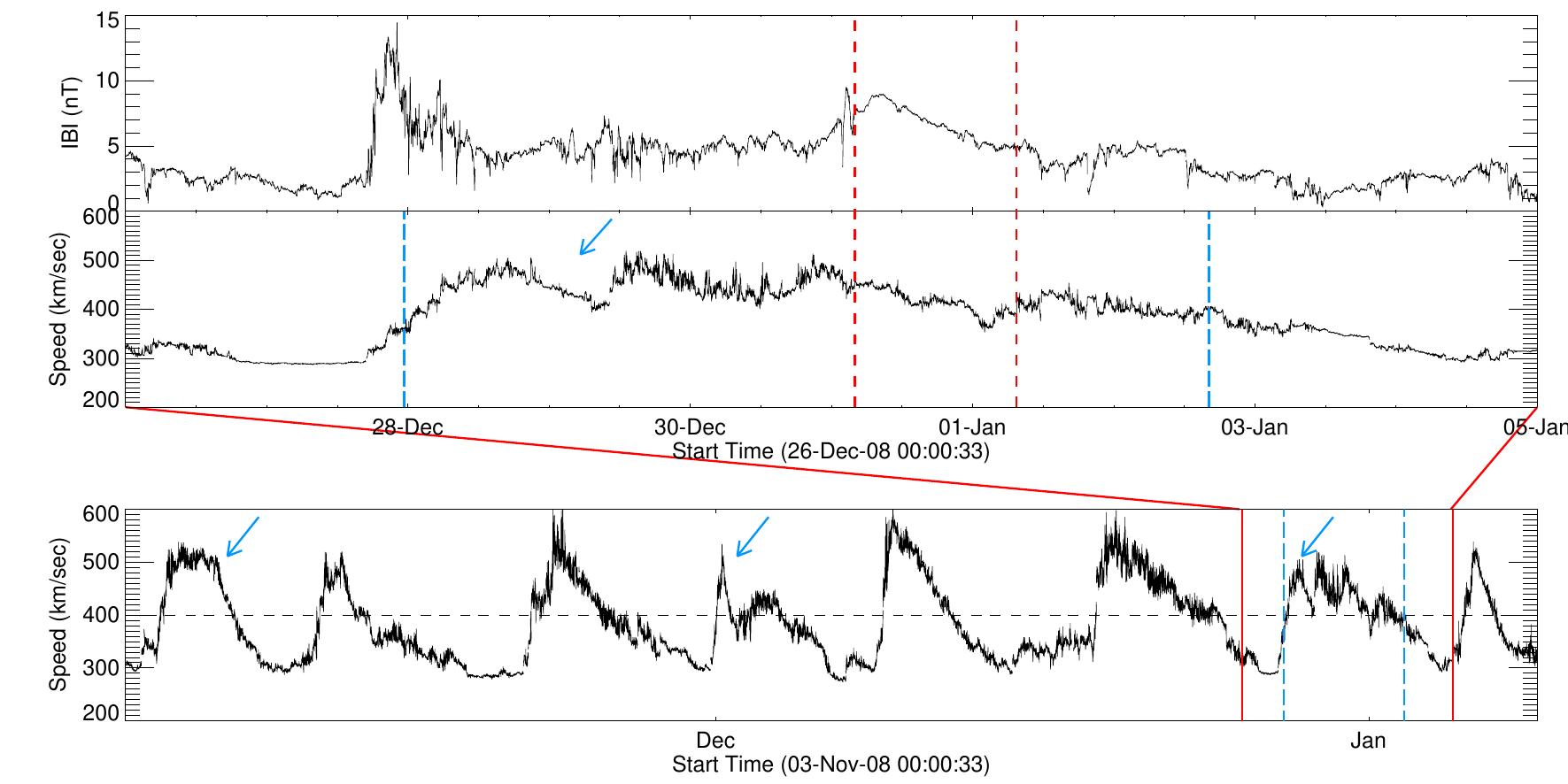}
\caption{The top and middle panel show the in-situ observations of the total magnetic field and the speed of the solar wind, respectively as obtained by STEREO-B during the ICME event 1. The two red dashed lines mark the boundary of the MC. The bottom panel is the same as the middle panel but plotted for a 3-month window. The two red solid lines in the bottom panel mark the temporal window within which the top and middle panels are plotted. The cyan dashed lines mark the approximate boundary of the HSS within which the ICME was embedded. The cyan arrows mark the appearance of the same co-rotating HSS in the previous solar rotations. \label{cir1}}
\end{center}
\end{figure*}

\section{Results and discussion}\label{results}
Using observationally constrained magnetic and geometric parameters for the CME FRs as estimated at 10 R$_S$, we run the INFROS model for each of the six events. We discuss the model results and compare them with the in-situ observations in separate subsections \ref{without_hss}, \ref{with_hss} and \ref{with_hds} for the ICMEs that undergo isolated evolution, interaction with high-speed streams (HSS) and high-density streams (HDS), respectively. All the magnetic vectors of the ICMEs are expressed in Spacecraft Equatorial (SCEQ) coordinates, where the z-axis aligns with the solar rotation axis, the y-axis indicates solar west, and the x-axis completes the right-handed system.

\subsection{Events associated with isolated ICME evolution} \label{without_hss}
ICMEs associated with events 3 \& 4 undergo isolated evolution in absence of any significant interaction with HSS, HDS or any other ICME. Figure \ref{ICME3_4} shows the comparison of the INFROS model results with the observed magnetic vectors of the ICMEs. 

The smooth rotation as observed in the in situ profile of Bz component in event 4 indicates that the associated flux-rope axis has a low inclination angle with respect to the ecliptic plane which is consistent with its near-Sun axis orientation obtained from remote-sensing observations. On the other hand, the in-situ magnetic vectors of the ICME in event 3 show rotation in By component, indicating a highly inclined flux-rope axis which is also in agreement with the near Sun observations as obtained from GCS tilt (see Table \ref{table2}). Therefore, the flux-rope orientation obtained at 10 R$_S$ is found to be approximately maintained throughout the rest of the propagation phase of the ICME.

The ensemble model results obtained from INFROS for Event-3 and -4 exhibit a spread in the predicted magnetic field profiles (depicted as grey shaded regions in Figure \ref{ICME3_4}) at the locations of both the inner (MESSENGER) and outer (VEX/STEREO-B) spacecraft. This spread accounts for the uncertainties associated with determining the input parameters. Within this uncertainty limit, Figure \ref{ICME3_4} shows that both the strength and orientation of the underlying magnetic FR structure of the ICMEs are well captured by INFROS at two well separated heliocentric distances. 

We observe that a considerable spread in the model results stems from the uncertainty involved in determining the impact distance, which depends on the direction of propagation and the tilt angle of the flux-rope. Comparing a set of our ensemble results (shown by the cyan dashed lines) that corresponds to the lowest impact distance at the outer spacecraft, we find a very good agreement with the observed magnetic profiles. Contextually, if the CME magnetic-axis passes through the spacecraft location, the impact distance will be minimum, resulting in a $B_x$ profile of zero throughout the CME passage  \citep{Sarkar_2020}. Indeed, the average observed $\left| B_x \right|$  profile at the outer spacecraft (VEX and STEREO-B) for both Event-3 and -4, exhibits lower value (0.1 and 0.2) relative to the total field-strength, indicating that the impact distance at the outer spacecraft were minimal for both the events. 

Importantly, the axial field strength (B$_0$) inside a self-similarly expanding flux-rope decreases with heliocentric distance (r) following the relation, $B_0 \propto \frac{1}{r^2}$ \citep{Sarkar_2020}, which is also one of the inherent assumption in INFROS. Therefore, the nice agreement between the INFROS model results and the in-situ observations obtained for events 3 and 4 reveals that the ICME evolution is self-similar with the field strength (B$_0$) decreasing as $\propto \frac{1}{r^2}$. Therefore, INFROS proves to be a useful forecasting tool to predict the magnetic vectors of isolated ICMEs.

Intuitively, the observed B-field data as obtained from the inner spacecraft may help us to constrain the model parameters, resulting in more accurate predictions of magnetic vectors at the outer spacecraft. In order to examine this hypothesis, we first select the ensemble model output (see the red dashed lines in the left panels of Figure \ref{ICME3_4}) that best fits the magnetic vectors observed by MESSENGER. We obtain this best fit by minimizing the normalised rms error ($\Delta_{rms}$) in modelled $B_y$, $B_z$ and total magnetic field strength $\left| B \right|$, where $\Delta_{rms}$ is defined as
$$\Delta_{rms}= \frac{1}{\left| B \right|^{obs}_{max}} \sqrt{ \frac{\sum^{N}_{i=1}(B_{model}(t_i)-B_{obs}(t_i))^2}{N}} $$ \citep{2019JGRA..124.4960G,2022pal}.

Here, $B_{obs}(t_i)$ and $B_{model}(t_i)$ represent the observed and modeled magnetic fields, respectively, at time $t_i$, sampled over N data points of the modeled magnetic field. ${\left| B \right|^{obs}_{max}}$ denotes the maximum intensity of the observed total magnetic field. 

Using the input parameters for the best fitted results at MESSENGER, the INFROS model results at the outer spacecraft is over-plotted as red dashed lines on the observed magnetic field data of VEX and STEREO-B for Event-3 and -4 respectively (see Figure \ref{ICME3_4}). These red dashed lines, obtained by constraining the model outputs at the inner spacecraft, exhibit remarkable agreement with the observed magnetic vectors at the outer spacecraft, significantly reducing the uncertainty in predicting the magnetic vectors at those locations. In examining the correlation between the observed and modelled Bz profile (red dashed line) at the outer spacecraft, Pearson correlation analysis reveals a significant positive correlation coefficient of r = 0.90 ($95\%$CI [0.85, 0.92], p $<$ 0.001) for Event-3 and r = 0.89 ($95\%$CI [0.88, 0.90], p $<$ 0.001) for Event-4. The confidence interval (CI) suggests that the true correlation coefficient is likely to fall between 0.85 (0.88) and 0.92 (0.90) for Event-3 (4) with 95\% confidence. Moreover, the associated $\Delta_{rms}$ in $B_z$ yields significantly lower value at 0.15 and 0.20 for Event-3 and 4 respectively.

This result holds significant importance in the realm of space weather forecasting, highlighting the significance of multi-spacecraft observations at various heliocentric distances below 1 AU. Such observations help in constraining space-weather forecasting models, thereby enhancing their accuracy before making predictions at 1 AU. Furthermore, the remarkable accuracy of INFROS model predictions at the outer spacecraft, after constraining the model parameters from the observations at the inner spacecraft, also shows prediction efficacy of INFROS for forecasting the magnetic vectors of ICMEs.

\begin{figure*}[t!]
\begin{center}
\includegraphics[width=.32\textwidth,clip=]{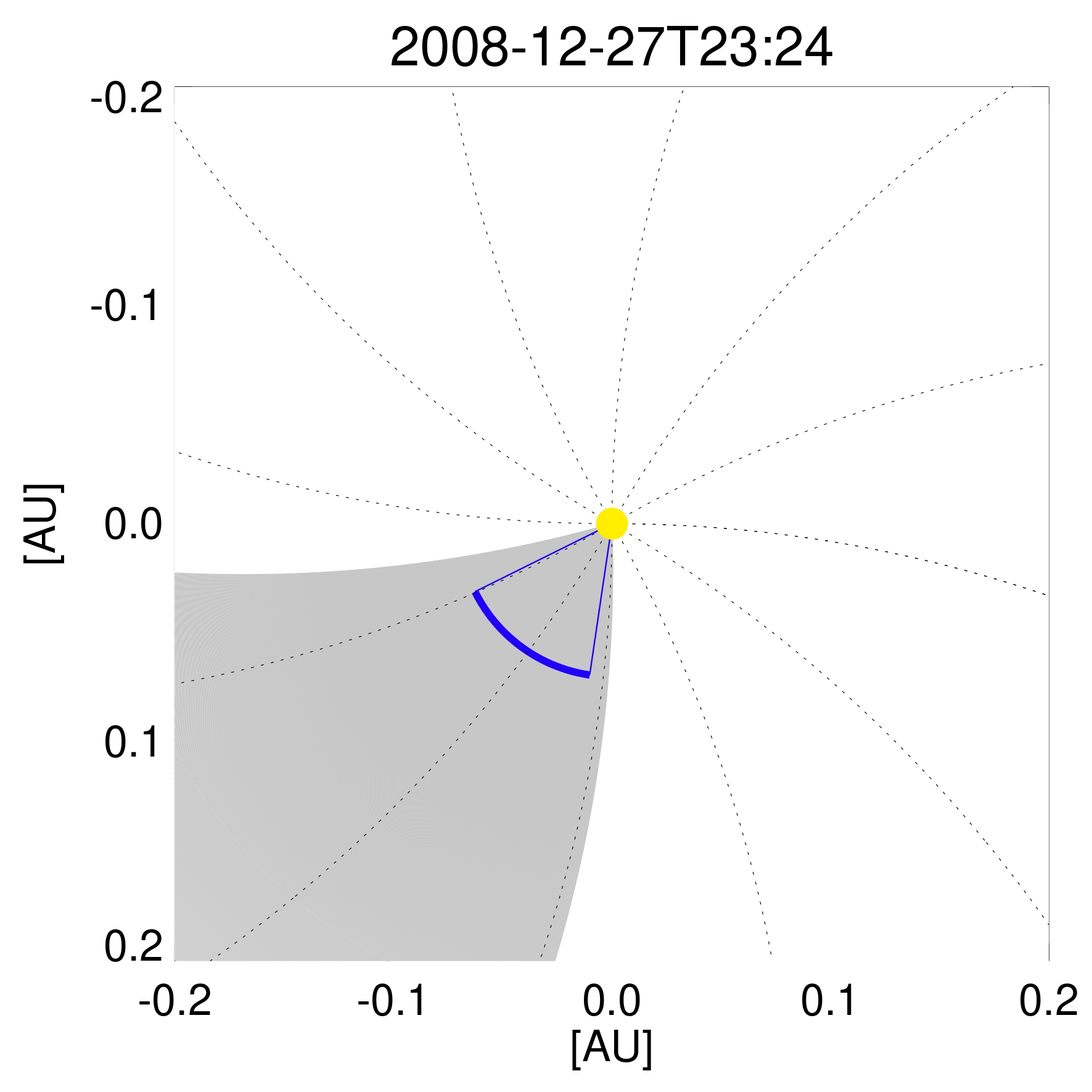}
\includegraphics[width=.32\textwidth,clip=]{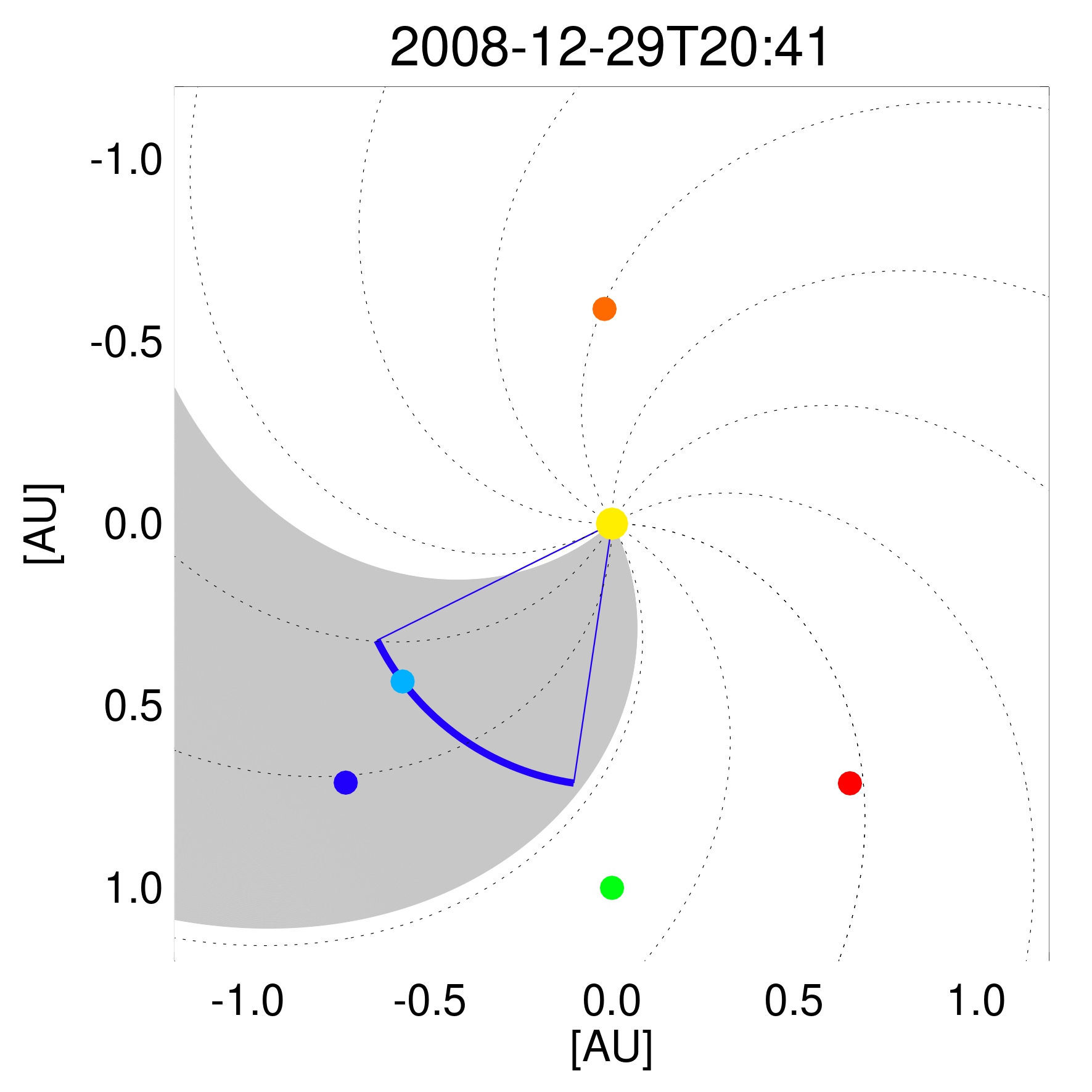}
\includegraphics[width=.32\textwidth,clip=]{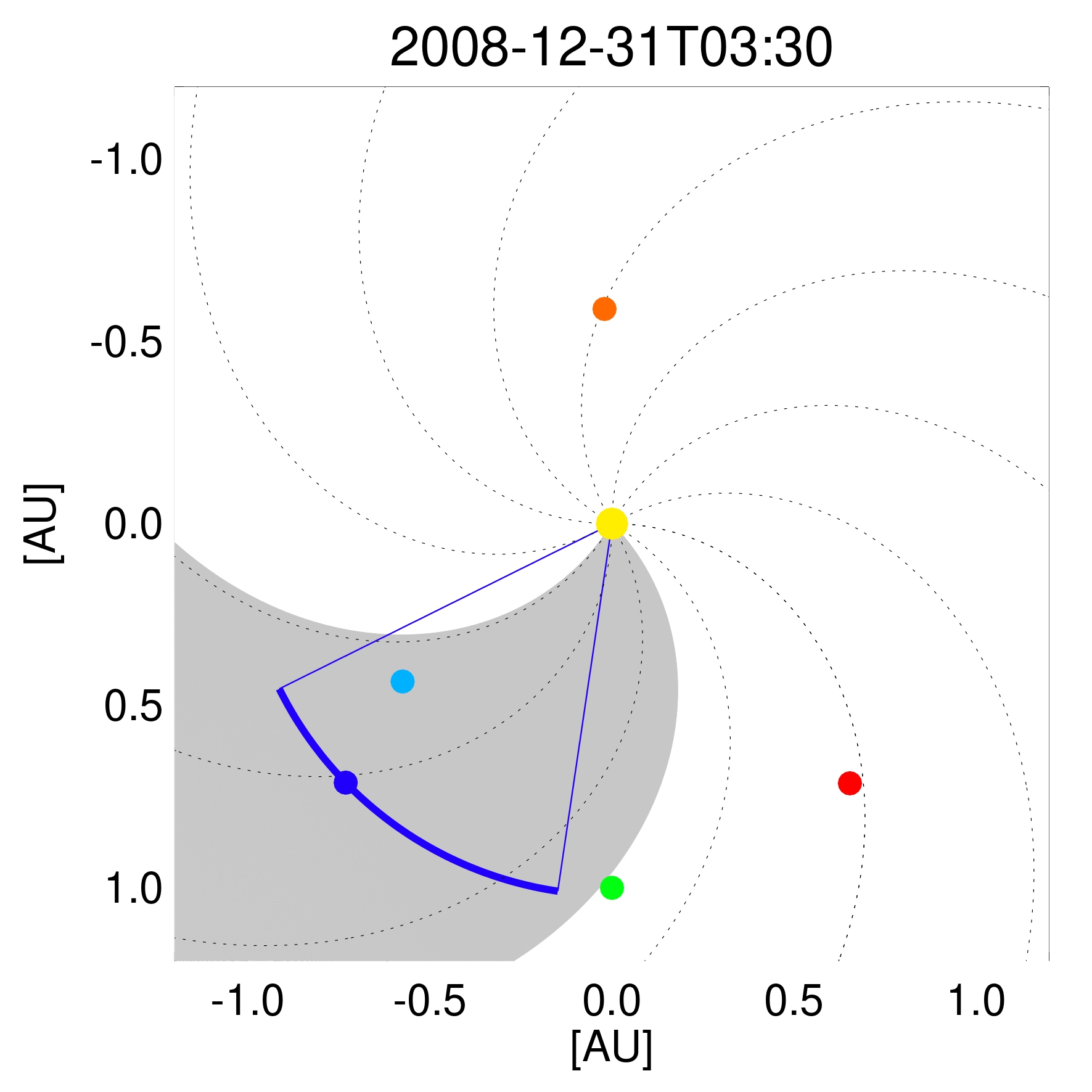}
\includegraphics[width=\textwidth,clip=]{planet_lebels.pdf}
\caption{Propagation of the CME leading edge and the associated HSS at different times of the CME evolution from the Sun to 1 AU as viewed on the ecliptic plane (Event 1). The black dotted lines denote the Parker's spiral. The gray shaded region shows the HSS associated with the CME. The curved blue lines depict the CME leading edge. The straight blue lines mark the axis of the two legs of the CME assuming the GCS geometry.\label{HSS_event1}}
\end{center}
\end{figure*}
\begin{figure*}[t!]
\begin{center}
\includegraphics[width=.9\textwidth,clip=]{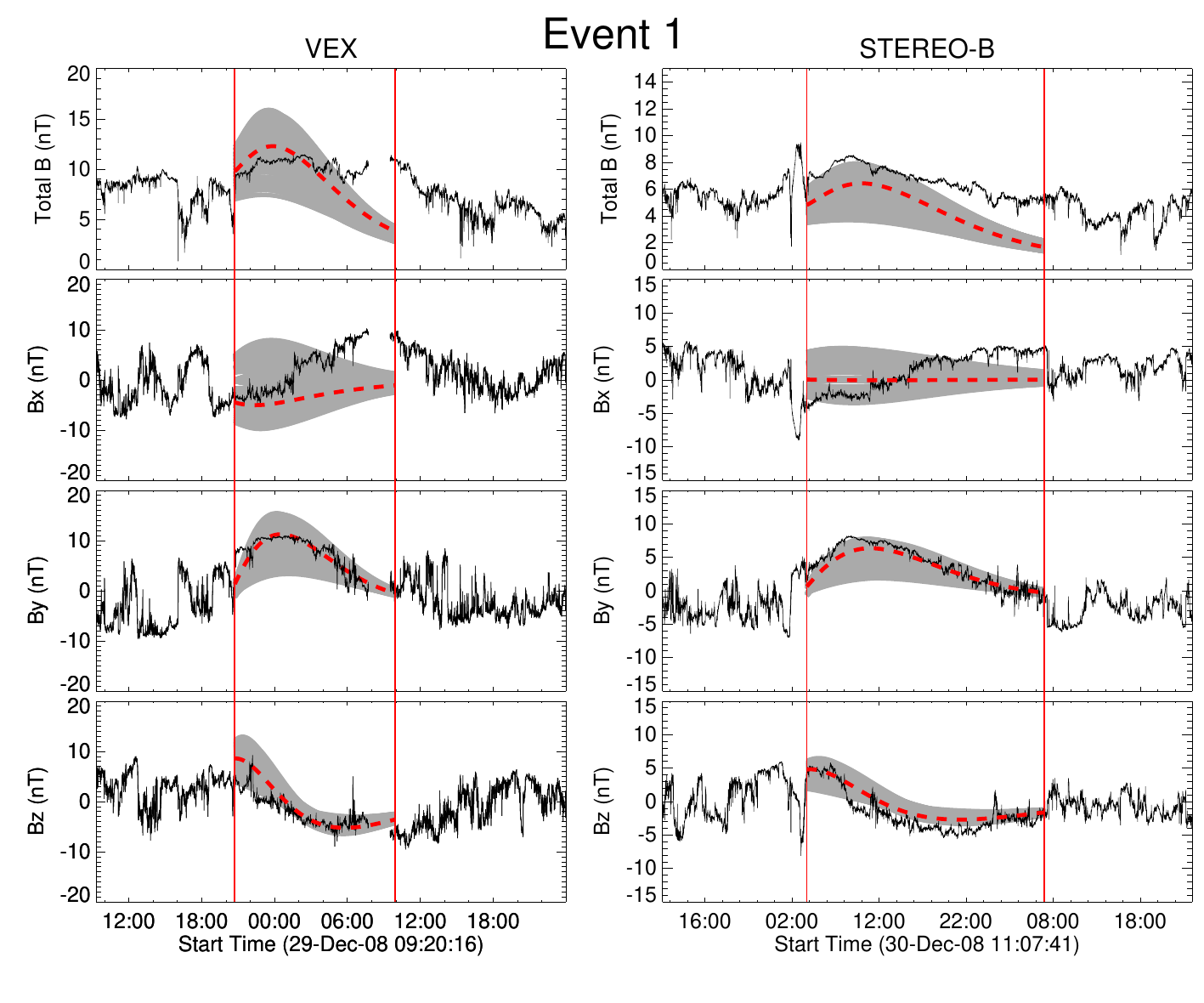}
\caption{The black solid lines within the boundary marked by the vertical red lines, denote the magnetic vectors (in SCEQ coordinates) of the ICME which is sequentially observed by VEX and STEREO-B for Event-1. The gray shaded regions are the results obtained from the model predictions. Amongst these ensemble model results, the one that best fits the magnetic vectors observed by VEX is denoted as a red dashed line in each plot.\label{ICME1_infros}}
\end{center}
\end{figure*}

\begin{figure*}[t!]
\begin{center}
\includegraphics[width=\textwidth,clip=]{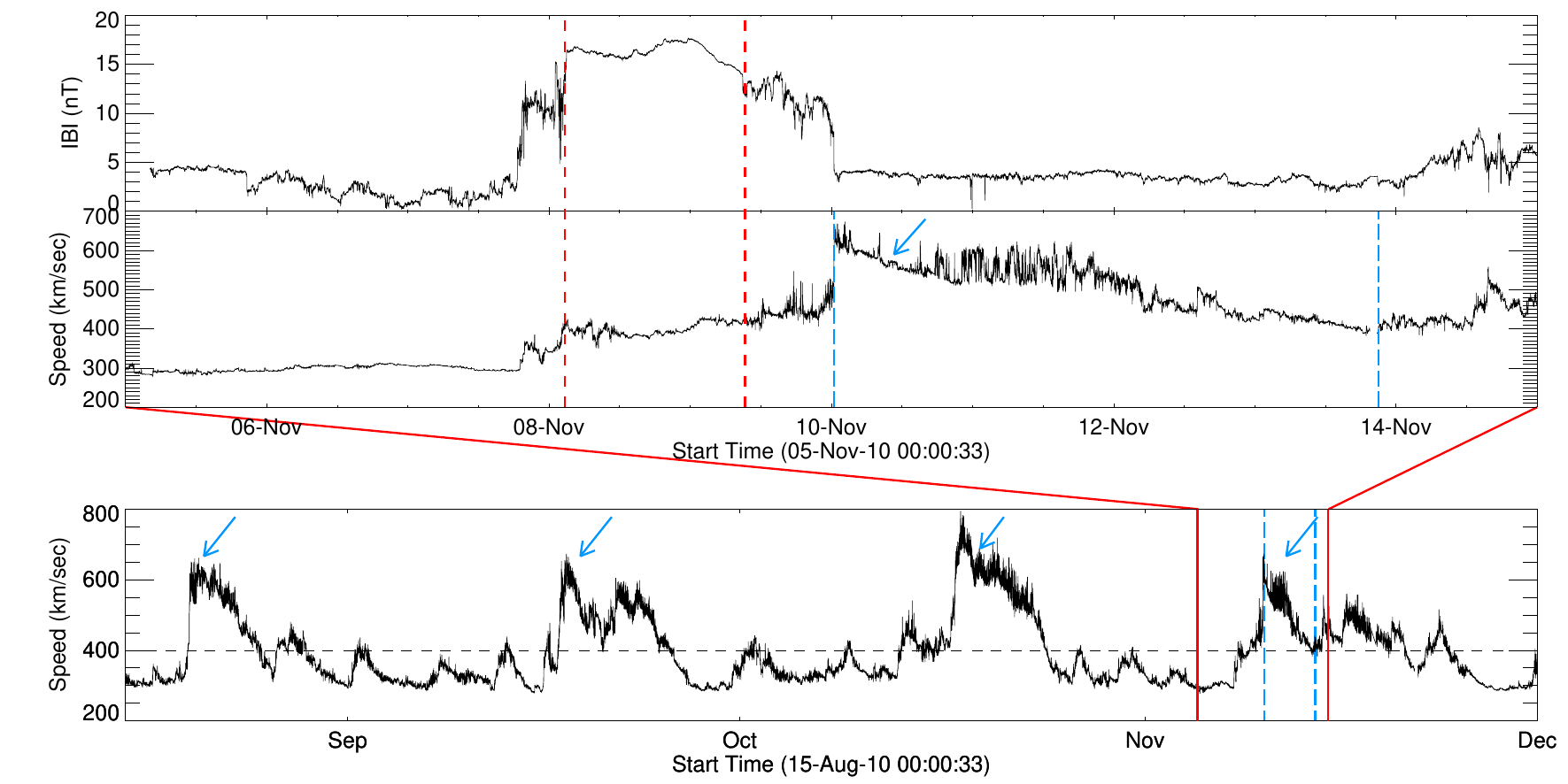}
\caption{The top and middle panels show the in-situ observations of the total magnetic field and the speed of the solar wind, respectively as obtained by STEREO-B during the ICME event 2. The two red dashed lines mark the boundary of the MC. The bottom panel is the same as the middle panel but plotted for a much wider temporal window of around four months. The two red solid lines in the bottom panel mark the temporal window within which the top and middle panels are plotted. The cyan dashed lines mark the approximate boundaries of the HSS followed by the ICME. The cyan arrows mark the appearance of the same co-rotating HSS in the previous solar rotations.}\label{insitu_event2}
\end{center}
\end{figure*}

\subsection{Events associated with interaction between ICME and HSS}\label{with_hss}
ICMEs associated with event no 1, 2 and 5 interact with the HSS during its propagation in interplanetary space. Out of these three events, event 2 and 5 were associated with cases where the ICME is followed by a HSS. On the other hand, the ICME in event 1 was embedded in the HSS during its entire propagation period from Sun to 1 AU. We describe the results of each event in separate subsections as below.

\subsubsection{ICME embedded in HSS}\label{event_1}

Figure \ref{cir1} shows the in-situ observations obtained from STEREO-B for event 1. The vertical cyan dashed lines drawn in the velocity plot of the solar wind (see middle panel) indicate that the leading edge of the HSS arrives at STEREO-B approximately on 28 December 2008 and the trailing edge passes through it towards the end of 2 January 2009. 
The boundaries of the HSS are selected based on the criteria that the solar wind speed inside the HSS should be greater than 400 $km \ s^{-1}$ (marked as horizontal black dashed line in the lower panel). We found that the HSS is co-rotating in nature and was also present in previous two solar rotations (indicated by the cyan arrows in lower panel). The MC boundaries as indicated by the two red dashed lines lie inside the temporal passage of the HSS, clearly indicating that the MC was embedded in the HSS when it arrives at 1 AU.

We further investigate the heliocentric distance at which the ICME starts to interact with the HSS during its propagation from the Sun to 1 AU. Assuming that the HSS co-rotates with the Sun, we reconstruct its morphological evolution in the ecliptic plane at different phases of the CME propagation (see Figure \ref{HSS_event1}). We first determine the time of arrival of the westernmost and easternmost boundary of the HSS at 1 AU as indicated by the cyan dashed lines in figure \ref{cir1}. Assuming that the HSS co-rotates following the synodic rotation speed 2.7$\times 10^{-6}$ rad s$^{-1}$, the time interval ($\approx$ 137 hours) between the arrival of the two boundaries of the HSS yields its angular width as $\approx$ 76$^{\circ}$. Finally, taking the average in-situ arrival speed of the HSS as 450 km s$^{-1}$, we use the Parker's spiral solution to connect the two boundaries of the HSS to the Sun as shown in Figure \ref{HSS_event1}. The three panels in  Figure \ref{HSS_event1} represent three different time instants that include when the CME leading edge height (projected on the ecliptic plane) is obtained from white-light observations by applying the GCS model (left panel), the leading edge of the magnetic cloud arrives at VEX (middle panel) and STEREO-B (right panel). Taking the arrival time of the HCS at STEREO-B as the reference time and using the synodic rotation speed, we de-rotate the HSS structure to those three different instants in order to find out where the CME starts to interact with the HSS.

%it is expected that the ICME would maintain almost similar expansion coefficient ($\kappa$) as obtained from the GCS   
%We first determine the temporal duration ($\approx$ 137 hours) between the leading and rear edge of the HSS as indicated by the cyan dashed lines in figure \ref{cir1}. This helps us to convert the temporal duration into the angular extent subtended by the HSS assuming that it rotates with solar rotational period of 2.7$\times 10^{-6}$ rad.sec$^{-1}$.  In order to 
 
%Although there is self-similarity, but the expansion is not self-similar

\begin{figure*}[t!]
\begin{center}
\includegraphics[width=.32\textwidth,clip=]{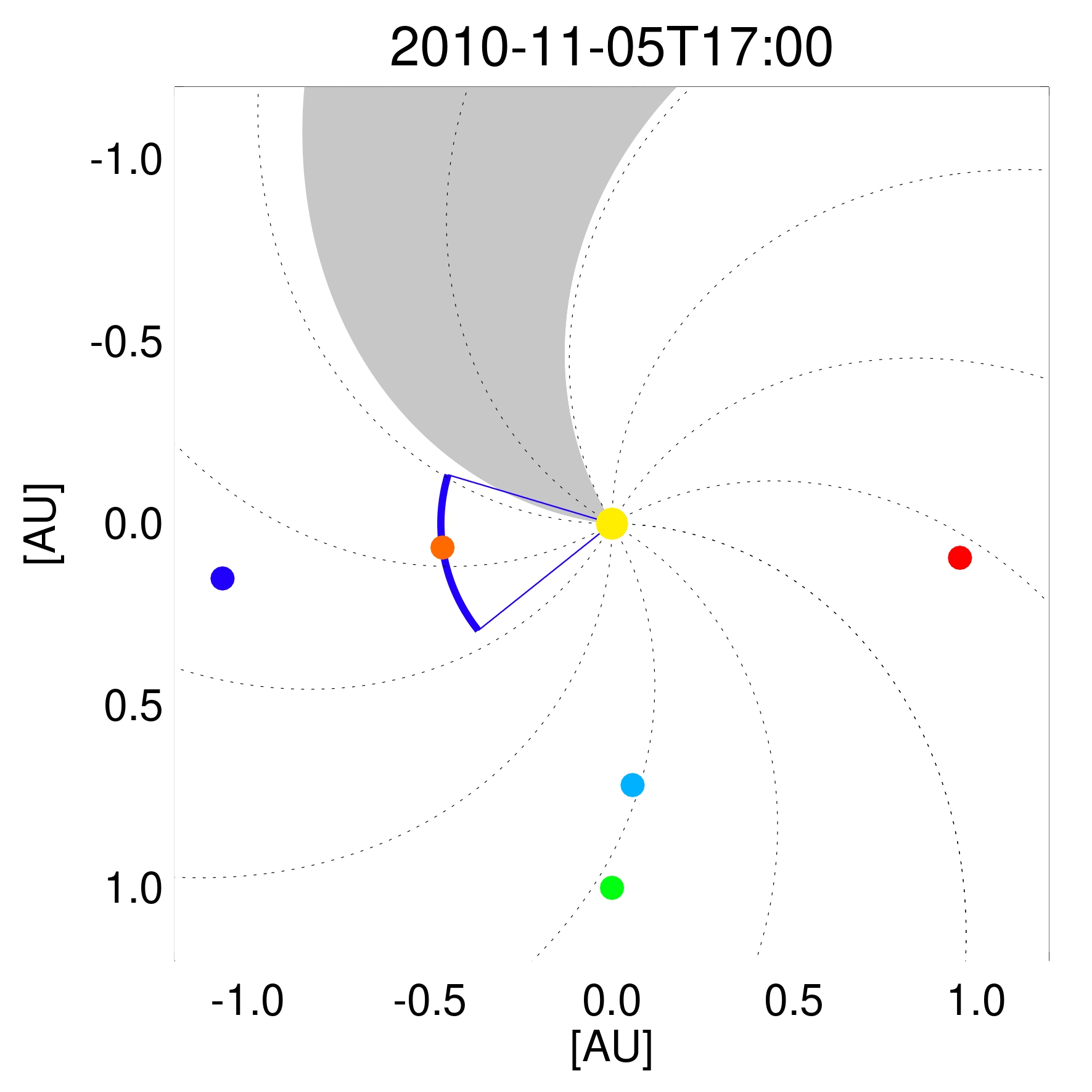}
\includegraphics[width=.32\textwidth,clip=]{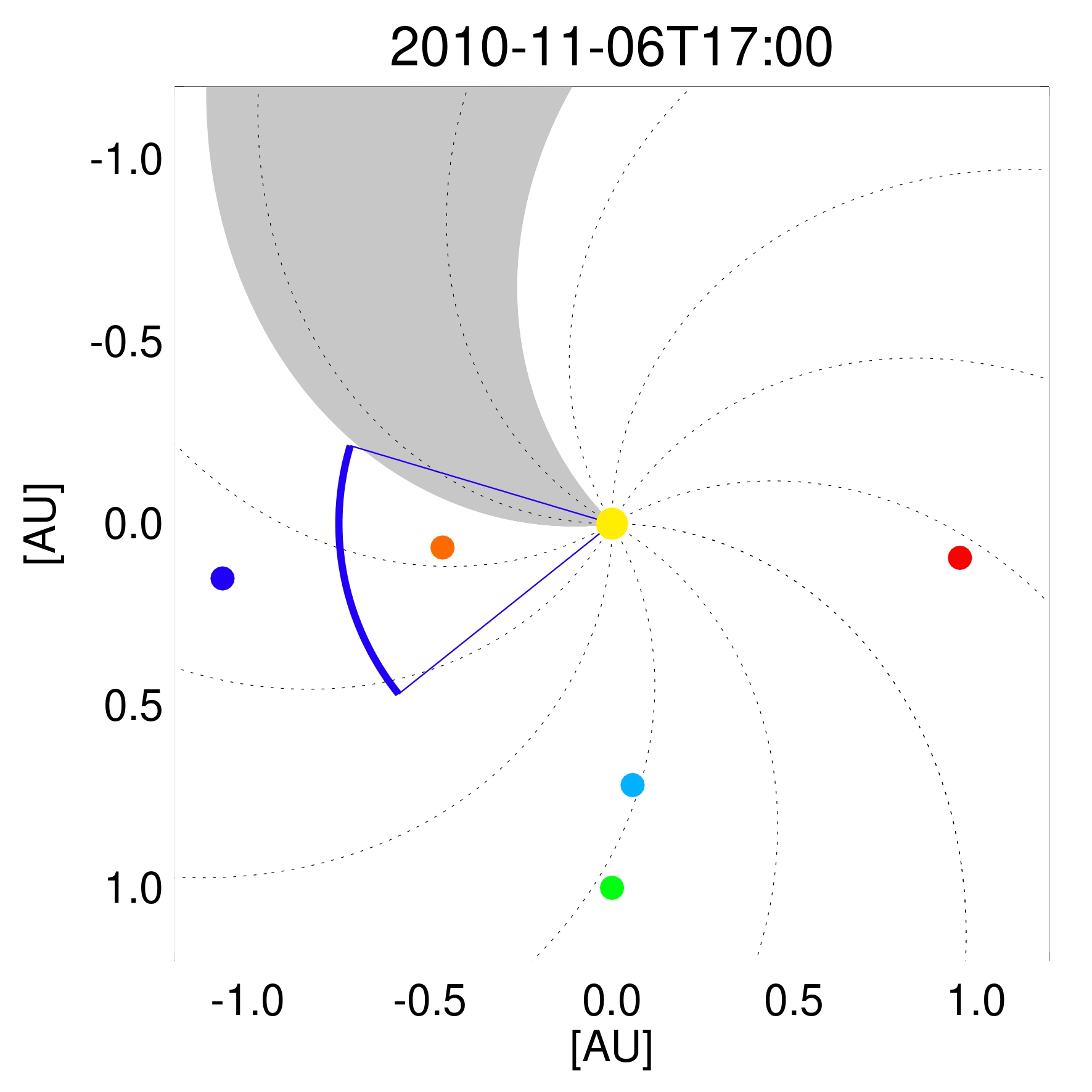}
\includegraphics[width=.32\textwidth,clip=]{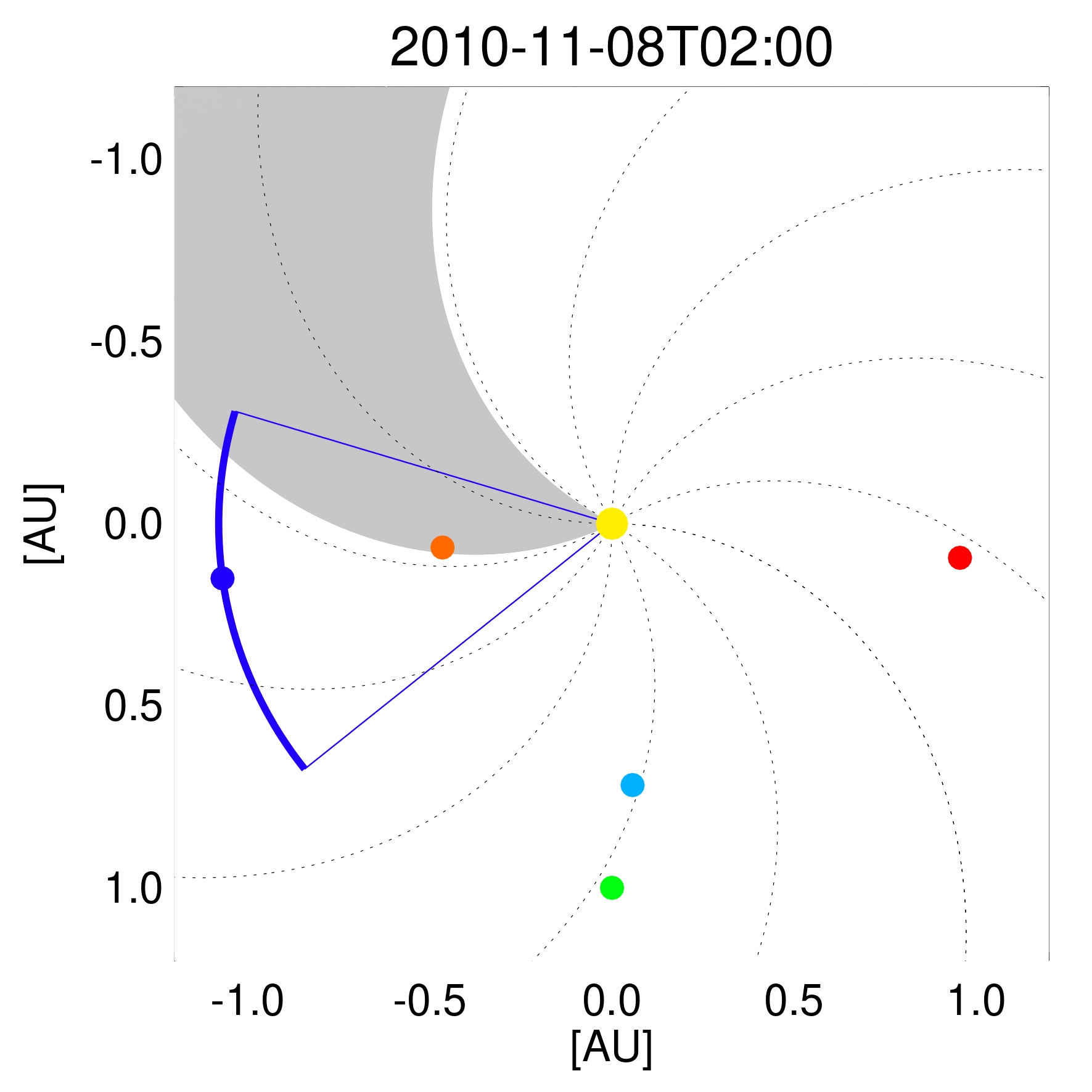}
\includegraphics[width=\textwidth,clip=]{planet_lebels.pdf}
\caption{Propagation of the CME leading edge and the associated HSS at different times of the CME evolution from the Sun to 1 AU as viewed on the ecliptic plane. The black dotted lines denote the Parker's spiral. The gray shaded region shows the HSS associated with the CME. The curved blue lines depict the CME leading edge. The straight blue lines mark the axis of the two legs of the CME assuming the GCS geometry. \label{ICME2_cir}}
\end{center}
\end{figure*}

Figure \ref{HSS_event1} clearly depicts that the ICME was embedded within the HSS throughout its propagation from the Sun to 1 AU. This implies that the background solar wind environment for the ICME remains almost similar during its whole propagation path. Therefore, as an interesting consequence, we find that the ICME evolution for this event is very much similar to the isolated evolution of an ICME, as the modelled magnetic field strength at both VEX and STEREO-B under the assumption of self-similar expansion are consistent with the observed in-situ values obtained at the respective spacecraft (see Figure \ref{ICME1_infros}). Along with the total field strength (B), the Bz and By components of the magnetic vectors are also well reproduced by the model at both the spacecraft within the uncertainty limit. Similar to events 3 and 4, constraining the ensemble model results from inner spacecraft (VEX) observation, we get significantly high correlation of r = 0.92 ($95\%$ CI [0.90, 0.93], p $<$ 0.001) and lower $\Delta_{rms}$ (=0.19), between the observed and modelled Bz profile (red dashed line) at STEREO-B. 

Notably, the observed Bx component of the MC at both the spacecraft shows rotation which may be only possible when the cross-sectional geometry of the flux-rope significantly deviates from the circular shape. As INFROS assumes a circular geometry for the flux-rope cross-section, the Bx component at both the spacecraft were not well captured by the model. However, consistency in the overall magnetic vectors and the field strength obtained from the observed and modelled profile of the MC indicates that if an ICME remains embedded in a HSS throughout its propagation and does not interact with any other ICME then it also follows the self-similar expansion.        

\subsubsection{ICME followed by a HSS}
\textbf{Event 2:} The in-situ observations of Event 2 obtained at STEREO-B clearly suggests the arrival of the MC followed by a HSS (see Figure \ref{insitu_event2}). The MC arrives at STEREO-B on 8 November 2010 at around 02:00 UT with a speed of $\approx$ 400 km s$^{-1}$. The speed of the MC within its temporal passage at STEREO-B as indicated by the two red dashed vertical lines in the middle panel of Figure \ref{insitu_event2} remains almost constant to $\approx$ 400 km s$^{-1}$. This indicates that the MC does not undergo any expansion during its passage through STEREO-B as its expansion is ceased by the following HSS that arrives at STEREO-B on 10 November 2010 at around 3:00 UT. Notably, the HSS was also present in the previous three solar rotations as shown in the bottom panel of Figure \ref{insitu_event2}.

\begin{figure*}[t!]
\begin{center}
\includegraphics[width=\textwidth,clip=]{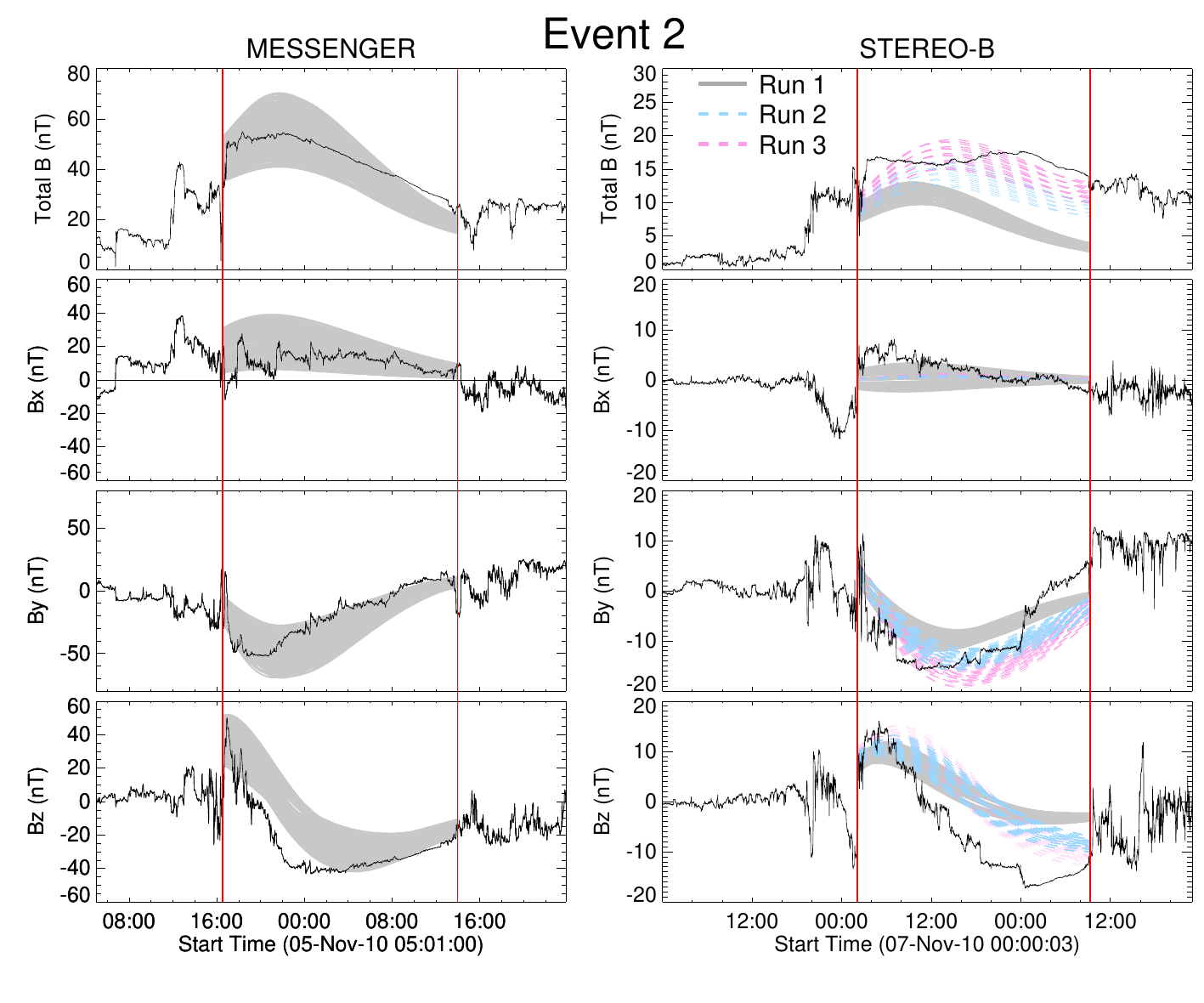}
\caption{The black solid lines within the boundary marked by the vertical red lines, denote the magnetic vectors (in SCEQ coordinates) of the ICME, which is sequentially observed by MESSENGER and STEREO-B for Event-2. The gray shaded regions are the model results obtained for Run1. The cyan and magenta coloured dashed lines in right panels denote the model results for Run2 and Run3, respectively \label{ICME2}}
\end{center}
\end{figure*}

\begin{figure*}[h!]
\begin{center}
\includegraphics[width=\textwidth,clip=]{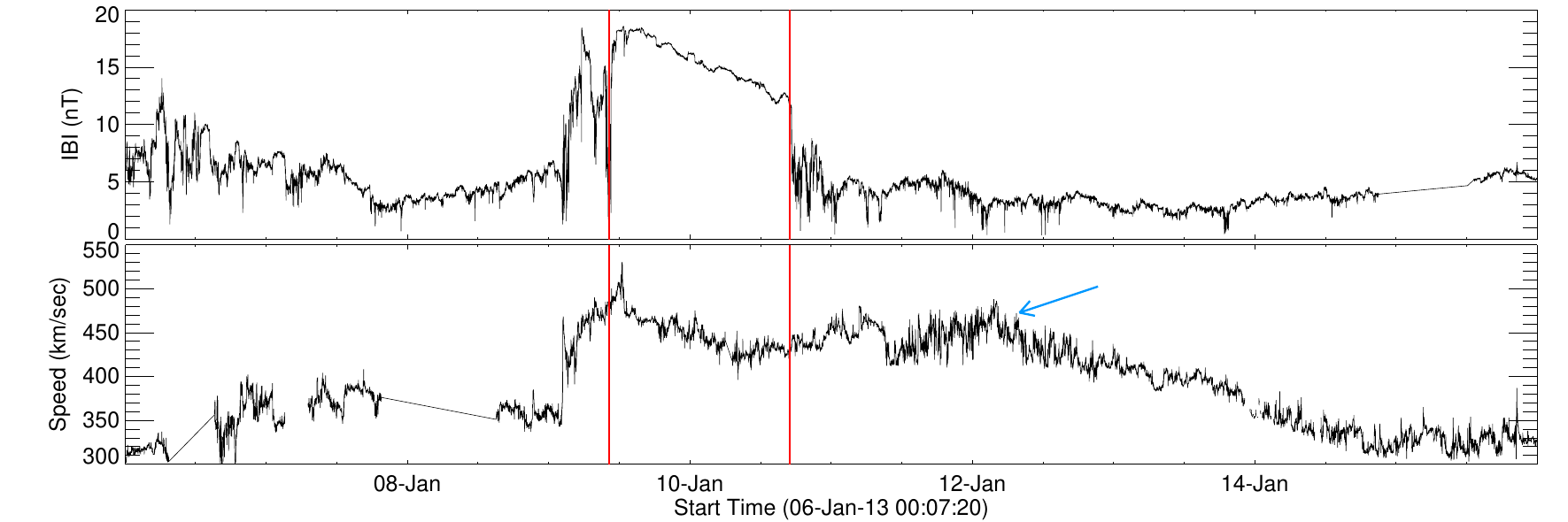}
\caption{The in-situ observations of the total magnetic field (top panel) and the speed (bottom panel) of the solar wind as obtained from STEREO-A during the ICME event 5. The two red solid lines mark the boundary of the MC. The cyan arrow mark the HSS following the MC.\label{hss5}}
\end{center}
\end{figure*}

\begin{figure*}[h!]
\begin{center}
\includegraphics[width=.8\textwidth,clip=]{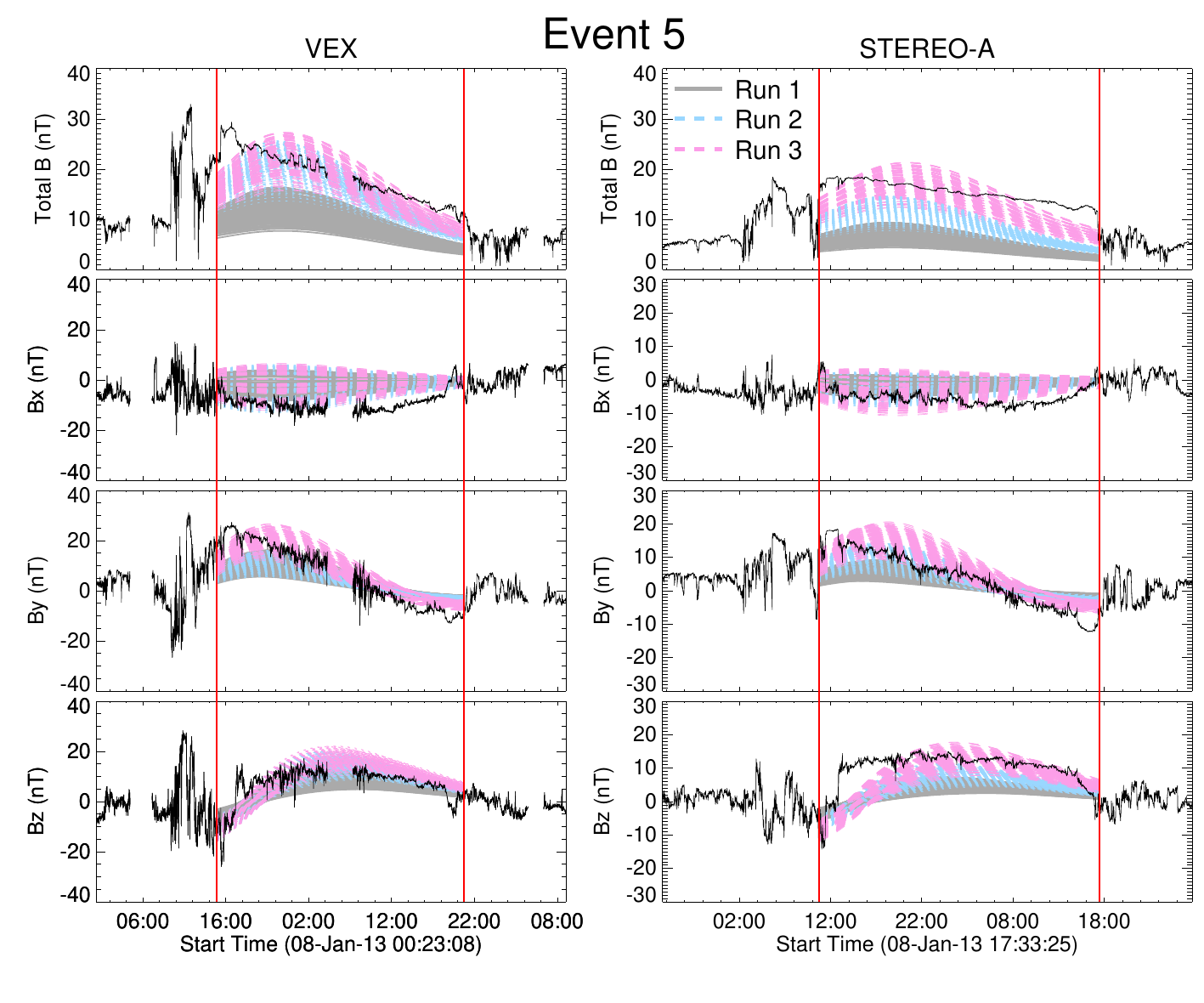}
\caption{ The black solid lines within the boundary marked by the vertical red lines, denote the magnetic vectors (in SCEQ coordinates) of the ICME, which is sequentially observed by VEX and STEREO-A for Event-5. The gray shaded regions are the model results obtained for Run1. The cyan and magenta coloured dashed lines in right panels denote the model results for Run2 and Run3 respectively \label{ICME5}}
\end{center}
\end{figure*}

\begin{figure*}[btp]
\begin{center}
\includegraphics[width=\textwidth,clip=]{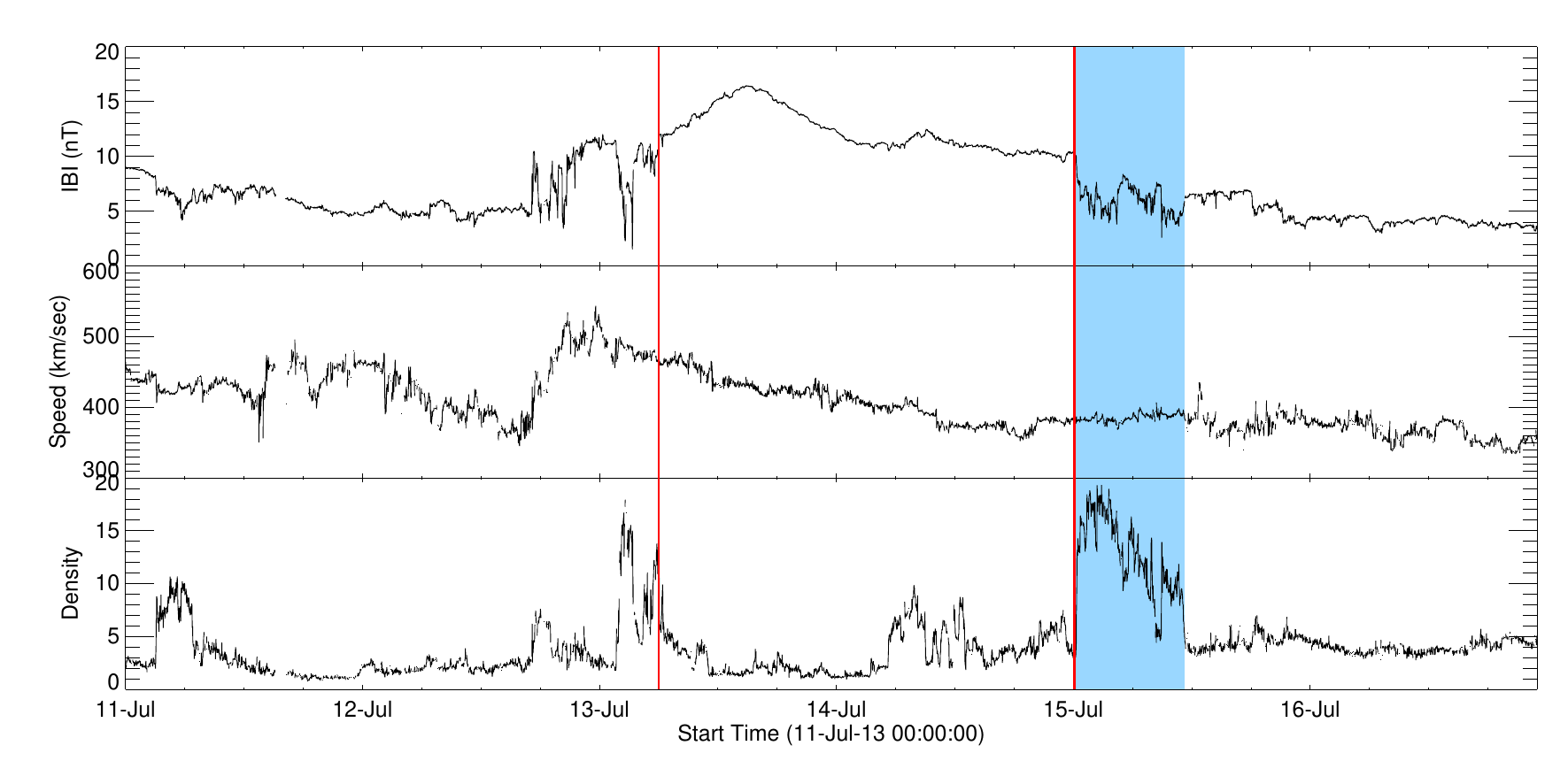}
\caption{The top, middle and bottom panels show the in-situ observations of the total magnetic field, the speed of the solar wind and the proton density, respectively as obtained by Wind during the ICME event 6. The two red dashed lines mark the boundary of the MC. The blue shaded region followed by the MC marks the location of the high density stream.\label{ICME6_insitu}}
\end{center}
\end{figure*}
 
\begin{figure*}[t!]
\begin{center}
\includegraphics[width=.8\textwidth,clip=]{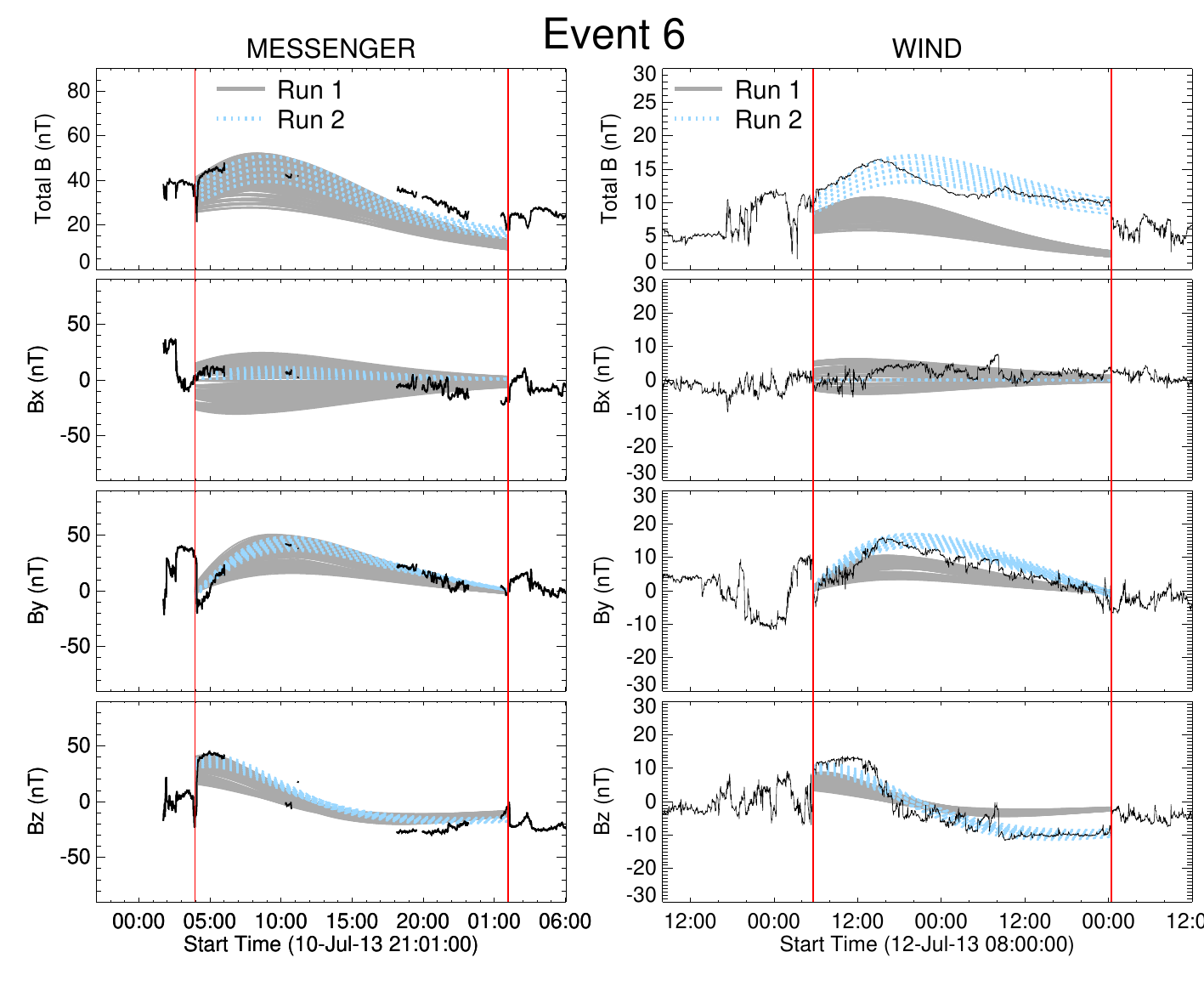}
\caption{The black solid lines within the boundary marked by the vertical red lines, denote the magnetic vectors (in SCEQ coordinates) of the ICME which is sequentially observed by MESSENGER and WIND for Event-6. The gray shaded regions are the model results obtained for Run1. The cyan and magenta coloured dashed lines in the right panels denote the model results for Run2 and Run3 respectively\label{ICME6}}
\end{center}
\end{figure*}

Following the same approach as discussed in the section \ref{event_1}, we identify the approximate heliocentric distance at which the ICME starts to interact with the HSS. The reconstructed locations of the HSS and the CME leading edge height at different time instants during its propagation from Sun to Mercury and STEREO-B, are illustrated in Figure \ref{ICME2_cir}. The left panel in Figure \ref{ICME2_cir} depicts the location of the HSS when the leading edge of the MC arrives at MESSENGER. It is clear from this qualitative analysis that there was no interaction between the ICME and the HSS during its propagation from Sun to MESSENGER. Therefore, the interaction must have occurred in between MESSENGER and STEREO-B. Using the drag based model \citep{DBM}, we evolve the leading edge of the CME at different time instants during its propagation from MESSENGER to STEREO-B and also track the location of the associated HSS at those time frames. This helps us identify the approximate timing and the heliocentric distance (0.7 AU) at which the western boundary of the HSS starts to interact with the eastern flank of the ICME as depicted in the middle panel of Figure \ref{ICME2_cir}. The right panel of the figure depicts the ongoing phase of the interaction when the MC leading edge arrives at STEREO-B. 

The INFROS model outputs for both the spacecraft under the assumption of self-similar expansion (Run1) are plotted as gray shaded region in Figure \ref{ICME2}. Notably, the modelled field strength and the orientation of the ICME magnetic vectors are in good agreement with the observed in situ values at MESSENGER. However, the INFROS model results significantly underestimate the observed field strength of the ICME detected at STEREO-B. This in-accuracy in estimating the field-strength at 1 AU, results in an underestimation of the Bz strength by $\approx$ 15 nT. 

Notably, the underestimation of the field strength at STEREO-B may occur if the impact distance, which refers to the distance between the FR axis and the spacecraft crossing path, is modelled as larger in Run1 as compared to the actual scenario. In such a case, the modelled spacecraft crossing path may miss the stronger field in the vicinity of the FR axis. On the other hand, the spacecraft would encounter the strongest field strength of the FR if it passes through the FR axis when the impact distance is zero. However, we notice that the ensemble model results from Run1 include the model prediction corresponding to zero impact distance at STEREO-B, eliminating the possibility of underestimating the field-strength due to large impact distance.

The consistency of the model results at MESSENGER indicates that the ICME exhibits self-similar expansion from Sun to Mercury and continues to evolve in that fashion until it interacts with the HSS during its propagation from Mercury to STEREO-B. Eventually, the HSS starts to interact with the ICME at approximately 0.7 AU, which does not allow the ICME to expand self-similarly anymore. Therefore, the evolution of the axial field strength (B$_0$) inside the ICME can be modelled following the relation, $B_0 \propto \frac{1}{r^2}$, only up to r= 0.7 AU. 

The underestimation of the modelled field strength at STEREO-B clearly depicts that the evolution of $B_0$ with heliocentric distance (r) above 0.7 AU is slower than the inverse square of r, i.e.  $B_0 \propto \frac{1}{r^m}; m < 2$. In order to incorporate this physics and see how the lower value of $m$ can better model the observed field strength of the ICME at STEREO-B, we perform a further run (Run2) with INFROS. In Run2, we consider the value of $m$ remains 2 up to the propagation path below 0.7 AU before the interaction starts and linearly decreases from 2 to 0  during the rest of the propagation path beyond 0.7 AU until the FR crosses the location of STEREO-B. The cyan dashed lines as shown in Figure \ref{ICME2}, depict the model results obtained from Run3. Notably, the results obtained in Run3 improves the modelled field strength as compared to that obtained from Run1 and Run2. 

However, the observed field strength is still underestimated in Run2, indicating that further lower value of $m$ may reproduce the observed magnetic vectors of the ICME at STEREO-B and can reveal its nature of expansion during the interaction phase with the HSS. Therefore, we perform another run (Run3) considering the value of $m$ to be 2 below 0.7 AU and linearly decreasing from 2 to -0.5 during the rest of the propagation beyond 0.7 AU until the FR crosses the location of STEREO-B. The choice of -0.5 as the lower end value of $m$ is made in order to get the best model result at STEREO-B after experimenting/iterating with several negative values as the lower limit of $m$. It is noteworthy that $m$ = 0 corresponds to the static evolution of an ICME without any expansion, whereas the negative values correspond to the compression of the ICME. Interestingly, the choice of $m$ in Run4 successfully reproduces the observed field strength at STEREO-B and also fairly models the strength of the Bz at the trailing edge of the MC (see Figure \ref{ICME2}). Importantly, the choice of $m$ in Run4 reveals that expansion of the ICME is not only ceased by the HSS, but the ICME also undergoes compression before it arrives at STEREO-B. Due to this compression, the total field strength as well as Bz  enhance relative to the expected values under self-similar expansion. This has important consequences for  space weather, as the enhancement in Bz at the tail end of such magnetic clouds of north-south polarity may lead to stronger geomagnetic storm if directed towards Earth \citep{fenrich_1998,kilpua_2012}. Notably, the assumption of self-similar expansion for the cases of non-isolated evolution of ICMEs would still be useful as it can at least be utilised to forecast the lower limit of the strength of Bz at 1 AU. 

\textbf{Event 5:} The ICME as detected by STEREO-A (Event 5) is also followed by a HSS similar to Event 2 (see Figure \ref{hss5}). The INFROS model results under the assumption of self-similar expansion (Run1) show that the model output underestimates the magnetic field strength at VEX and STEREO-A (see the gray shaded plots in Figure \ref{ICME5}). Notably, the ensemble model results (grey shaded plots) for Run1 include the output for zero impact distance of the FR at both VEX and STEREO-A. This excludes the scenario where the spacecraft passes away from the FR axis, resulting in missing the higher field strength region of the FR. Therefore, the underestimation of the field strength at both spacecraft in Run1 may occur under either of the two following circumstances: (i) Self-similar expansion is followed but the axial field strength ($B_{\circ}$=32$\pm$5 mG) used as input is underestimated; (ii) The ICME starts to interact with the HSS well before it arrives at VEX and therefore its expansion no longer remains self-similar. Exploring the above mentioned first scenario, we perform Run2 where we still assume self-similar expansion, but use a different $B_{\circ}$ value (53$\pm$5 mG) so as to get the correct field strength at VEX. However, with this modified $B_{\circ}$, we notice that the model at STEREO-A still underestimates the observed field strength (see Figure \ref{hss5}). This confirms that the ICME does not exhibit self-similar expansion, but undergoes compression exerted by the following HSS. Using a similar method as applied for Event 2, we find that the interaction between the ICME and HSS starts close to the Sun ($\approx$ 0.1 AU). This supports the aforementioned second scenario behind the discrepancy between the observations and model results. 

Finally, we show the results from Run3 with $m$ linearly decreasing from 2 to 1.6 starting from 0.1 AU to the rest of the propagation path until the FR crosses the location of STEREO-A. The Run3 model outputs are in remarkably good agreement with the observed field strength at both VEX and STEREO-A. Notably, the range of $m$ (2 to 1.6) used in Run3 for Event 5 is smaller than that (2 to -0.5) for Event 2. This indicates that the ICME in Event 5 undergoes lesser compression than Event 2 does. This might be due to the different relative speed between the HSS and the ICME in the two events. Notably, the speed difference between the ICME and the HSS in Event 2 was ($\approx$) 250 km s$^{-1}$ when they were detected by STEREO-B. However, their arrival speed at STEREO-A in Event 5 is almost comparable (see Figure \ref{hss5}). Therefore, the higher relative speed between the HSS and the ICME in Event 5 might have resulted in higher compression. %force exerted on the ICME by the following HSS.      

 %As the two spacecraft were radially aligned, it is expected that a same part of the ICME would be detected by the two spacecraft.

\subsection{ICME events associated with HDS interaction}\label{with_hds}

The in-situ observations of Event-6 as depicted in Figure \ref{ICME6_insitu} illustrate the arrival of the MC at Wind followed by a high density stream (HDS). Applying the similar analysis technique as discussed in section \ref{event_1}, we find that the HDS approximately starts to interact with the ICME at 0.5 AU. The INFROS model results under the assumption of self-similar expansion (Run1) are shown by the gray shaded region in Figure \ref{ICME6}. Although the model results of Run1 are in good agreement with the observed value at MESSENGER, the model underestimates the magnetic field strength  at 1 AU similar to Event 2 as discussed in previous section. This indicates that an HDS followed by an ICME plays similar role as HSS that does not allow the ICME to expand self-similarly during the interaction phase. As a result, the ICME field strength and Bz increase when it arrives at 1 AU.

Similar to Event-5, the ensemble model results of Run1 for Event-6 include the output for zero impact distance of the FR at both spacecraft, thereby eliminating the possibility of missing higher field strength due to the spacecraft crossing away from the FR axis. Considering that $m$ remains at 2 up to the propagation path below 0.5 AU before the interaction starts and linearly decreases from 2 to 0 during the rest of the propagation path, we perform another run (Run2) for this event. The results obtained from Run2 show remarkably good agreement with the observed values at both MESSENGER and Wind. This indicates the role of HDS to modify the nature of ICME expansion relative to the self-similar situation.

Previous studies have also suggested that the dependence of the axial field strength ($B_0$) of ICMEs on heliocentric distance ($r$) follows a power-law behavior ($B_0 \propto {r^{-m}}$). However, the above studies reported a range of power-law indices ($m$), such as $r^{-1.52}$ \citep{Wang_2005}, $r^{-1.75}$ \citep{Farrugia_2005}, $r^{-1.85}$ \citep{Gulisano_2005} , $r^{-1.64}$ \citep{Leitner_2007}, $r^{-1.95}$ \citep{Winslow_2015} and $r^{-1.6}$ \citep{spiros}. The broad range of $m$ obtained in the aforementioned studies likely stems from the fact that a clear distinction among the CMEs based on their background solar wind condition was not made. Instead, those studies relied on statistical fitting to derive an average power law index. Our results show that the value of m becomes less than 2 in cases where the evolution of the ICME no longer remains self-similar, but instead undergoes compression due to interaction with the background wind.

%However, our study highlights that the power law indices largely depend on whether an ICME is associated with isolated evolution or undergoes interaction.}

\section{Conclusion}\label{summary}
This study utilizes the state-of-the-art analytical model INFROS to investigate the evolution of six interplanetary coronal mass ejections (ICMEs). The analyzed events encompass the isolated interplanetary evolution of ICMEs as well as their interactions with high-speed streams (HSS) and high-density streams (HDS). By considering these diverse scenarios, we aim to gain a comprehensive understanding of the complex interplay between CMEs and their surrounding solar wind environment. In particular, these ICMEs were observed in sequence by two radially aligned spacecraft positioned at varying distances from the Sun, providing an unique opportunity to study their radial evolution from the perspective of space weather forecasting. By analyzing the corresponding CMEs and their source locations on the Sun, we determined the intrinsic flux-rope properties of each ICME near the Sun. To accomplish this, we utilized multi-wavelength remote sensing observations from spacecraft positioned at multi vantage points, including STEREO, SOHO, and SDO. The near-Sun flux-rope properties obtained from the observations were then used as input to INFROS in order to model the ICME evolution at different heliocentric distances. By comparing the model outputs with the in situ observations of the ICMEs obtained at the radially aligned spacecraft, such as MESSENGER, VEX, STEREO and Wind, we draw the following conclusions from our study.
\begin{itemize}
    \item INFROS model results are in very good agreement with the observed in-situ magnetic vectors of the ICMEs that undergo isolated evolution.

    %\item Isolated ICMEs follow self-similar expansion with the axial field strength ($B_0$) satisfying the relation  $B_0 \propto \frac{1}{r^2}$.  
    
    \item ICMEs embedded in high-speed streams also exhibit self-similar expansion as it evolves in a similar background solar wind environment throughout its propagation path and therefore INFROS model can be used to predict the magnetic vectors of an ICME under such circumstances.

    \item A considerable spread in the ensemble model results is observed, stemming from the uncertainty involved in determining the initial input parameters of the model based on near-Sun observations. The resultant uncertainty could be larger at outer spacecraft or L1 in the absence of three coronagraph viewpoints. However, the prediction uncertainty can be significantly reduced by constraining the model parameters using observations from inner spacecraft. For the three ICME events exhibiting self-similar expansion, INFROS model results achieve a 90\% correlation between observed and predicted Bz profiles, when the model parameters are constrained using inner spacecraft observations. This emphasizes the importance of multi-spacecraft observations and highlights the efficacy of INFROS for forecasting ICME magnetic vectors.

    \item Our study highlights that the power law index (m) related to the dependence ($B_0 \propto {r^{-m}}$) of the axial field strength ($B_0$) of ICMEs on heliocentric distance ($r$), 
    largely depends on whether an ICME is associated with isolated evolution or undergoes interaction. Distinguishing the isolated and interacting cases among the six events under study, we have found that isolated ICMEs follow self-similar expansion with $B_0$ satisfying the relation  $B_0 \propto {r^{-2}}$. On the other hand, $B_0$  decreases slowly with $r$ ($m < 2$) as compared to the inverse square of $r$ for the cases of interacting events.
    
    \item The assumption of self-similar expansion underestimates the field-strength for the interacting cases. Importantly, it provides a lower limit to it. 
    
    \item Due to the interaction of the ICME with high-speed streams and high-density streams, the total field strength of the ICME at 1 AU may significantly get enhanced as compared to the self-similar expansion situation. Such enhancement of the field strength at the trailing edge of magnetic clouds with north-south polarities may result in a decrease of the negative Bz value by up to 15 nT at 1 AU, as studied in this work. A decrease of 15 nT in $B_z$ for an Earth directed ICME with average speed ($v$) of 650 $km\ s^{-1}$ can be associated with a decrease of peak Dst index by 115 nT according to the empirical relationship, $\Delta~$Dst$~=~-0.017\times~V\times~\Delta B_z$ \citep{gopalswamy_dst}. This highlights the role of interactions leading to adverse space-weather.

\end{itemize}

As a future work, we plan to extend this study by modelling these events with the framework of global MHD models. This would allow us for a more detailed exploration of the interaction between ICMEs and the solar wind
from the perspective of space-weather forecasting.   

We thank the referee for helpful comments that improved the quality of this manuscript. RS acknowledges support from the project EFESIS (Exploring the Formation, Evolution and Space-weather Impact of Sheath-regions), under the Academy of Finland Grant 350015. NS and NG acknowledge support from the Indo-U.S. Science and Technology Forum (IUSSTF) Virtual Network Center
project (Ref. No. IUSSTF/JC-113/2019). EK acknowledges the ERC under the European Union's Horizon 2020 Research and Innovation Programme Project 724391 (SolMAG). The work at the University of Helsinki was performed under the umbrella of Finnish Centre of Excellence in Research of Sustainable Space (Academy of Finland Grant no. 312390, 336807). Open access is funded by Helsinki University Library.

\end{document}